\documentclass[10pt,letterpaper,aps,pra,twocolumn,superscriptaddress,floatfix]{revtex4-1}
\usepackage[latin1]{inputenc}
\usepackage{natbib}
\usepackage[latin1]{inputenc}
\usepackage{amsmath}
\usepackage{amsfonts}
\usepackage{amssymb}
\usepackage{bm}
\usepackage[pdftex]{graphicx}
\usepackage[colorinlistoftodos]{todonotes}
\usepackage{color}
\usepackage{MnSymbol}
\usepackage{enumerate}
\usepackage{bbold}
\usepackage{lipsum}
\usepackage{array}
\usepackage{tikz}
\usetikzlibrary{shapes,arrows}
\usetikzlibrary{positioning,arrows}
\usetikzlibrary{decorations.pathmorphing}
\usetikzlibrary{decorations.markings}

\begin{document}
\tikzset{
particle/.style={thick,draw=black, postaction={decorate},decoration={markings,mark=at position .5 with {\arrow[draw]{<}}}},
particle2/.style={thick,draw=black, postaction={decorate},decoration={markings,mark=at position .5 with {\arrow[draw]{<}}}},
gluon/.style={thick,decorate, draw=black,decoration={coil,aspect=0,segment length=3pt,amplitude=1pt}}
 }
\title{Anatomy of Fluorescence: Quantum trajectory statistics from continuously measuring spontaneous emission}
\author{Andrew N. Jordan}
\affiliation{Department of Physics and Astronomy, University of Rochester, Rochester, New York 14627, USA}
\affiliation{Center for Coherence and Quantum Optics, University of Rochester, Rochester, New York 14627, USA}
\affiliation{Institute for Quantum Studies, Chapman University, 1 University Drive, Orange, CA 92866, USA}
\author{Areeya Chantasri}
\affiliation{Department of Physics and Astronomy, University of Rochester, Rochester, New York 14627, USA}
\affiliation{Center for Coherence and Quantum Optics, University of Rochester, Rochester, New York 14627, USA}
\author{Pierre Rouchon}
\affiliation{Centre Automatique et Syst\`emes, Mines-ParisTech,
PSL Reseach University, 60, bd Saint-Michel, 75006 Paris, France}
\author{Benjamin Huard}
\affiliation{Laboratoire Pierre Aigrain, Ecole Normale Sup\'erieure-PSL Research University,
CNRS, Universit\'e Pierre et Marie Curie-Sorbonne Universit\'es, Universit\'e Paris Diderot-Sorbonne Paris Cit\'
e, 24 rue Lhomond, 75231 Paris Cedex 05, France}
\date{\today}
\begin{abstract}
We investigate the continuous quantum measurement of a superconducting qubit undergoing fluorescence.
The fluorescence of the qubit is detected via a phase-preserving heterodyne measurement, giving the fluorescence quadrature signals as two continuous qubit readout results.
By using the stochastic path integral approach to the measurement physics, we derive most likely paths between boundary conditions on the state, and compute approximate time correlation functions between all stochastic variables via diagrammatic perturbation theory.   We focus on paths that increase in energy during the continuous measurement.  Our results are compared to Monte Carlo numerical simulation of the trajectories, and we find close agreement between direct simulation and theory.  We generalize this analysis to arbitrary diffusive quantum systems that are continuously monitored.
\end{abstract}

\newcommand{\op}[1]{\hat{ #1}}                
\newcommand{\ket}[1]{\lvert#1\rangle}
\newcommand{\bra}[1]{\langle#1\rvert}
\newcommand{\pr}[1]{\ket{#1}\bra{#1}}
\newcommand{\ipr}[2]{\langle #1 | #2 \rangle}
\newcommand{\mean}[1]{\left\langle #1 \right\rangle}
\newcommand{\cw}{\circlearrowright}
\newcommand{\ccw}{\circlearrowleft}
\newcommand{\be}{\begin{equation}}
\newcommand{\ee}{\end{equation}}
\newcommand{\ra}{\rangle}
\newcommand{\la}{\langle}
\newcommand{\dd}{\mathrm{d}}

\newcommand{\dv}[2]{{\frac{\partial #1}{\partial #2}}}
\newcommand{\dvv}[2]{{\frac{\partial^2 #1}{\partial {#2}^2}}}
\newcommand{\EE}[1]{\text{\Large $\mathbb{E}$}\left(#1\right)}
\newcommand{\PP}[1]{\text{\Large $\mathbb{P}$}\left(#1\right)}
\newcommand{\dotex}{\frac{d}{dt}}
\newcommand{\tr}[1]{\text{Tr}\left(#1\right)}
\newcommand{\trr}[1]{\text{Tr}^2\left(#1\right)}
\newcommand{\trp}[2]{\text{Tr}_{#1}\left(#2\right)}
\newcommand{\RR}{{\mathbb R}}
\newcommand{\br}{\boldsymbol{r}}

\maketitle
\section{Introduction}
Fluorescence concerns the energy relaxation of a quantum system, emitting coherent light at its transition frequency. By continuously measuring the emitted light, it is possible to infer the evolution of the quantum state of the system in time -- its quantum trajectory~\cite{wiseman2009quantum,carmichael2009open}. In the simplest case of a qubit, a single photon is emitted when the qubit transitions from its excited to its ground state. If a single photon detector were used, the measurement would yield a discrete event of 0 or 1 photons emitted, collapsing the qubit state to its ground state (if a photon were detected), or partially collapsing the qubit to a shifted coherent superposition (if a photon is not detected)~\cite{PhysRevLett.68.580,Carmichael1989,Dum1992,Wiseman1993}.
In contrast, using a phase preserving amplifier \cite{caves1982quantum,hatridge2013quantum}
followed by heterodyne measurement of both quadratures of the fluorescence light mode~\cite{Wiseman1993,wiseman2009quantum,carmichael2009open,Barchielli2009} leads to diffusive stochastic quantum trajectories of the qubit state (see Fig.~1). Such trajectories were recently observed in a superconducting qubit by monitoring its fluorescence using a superconducting phase preserving amplifier~\cite{Campagne2015,six2015parameter}.

The feasibility of measuring diffusive quantum trajectories in superconducting qubits~\cite{murch2013observing,weber2014mapping} is an important development in this field, and motivates a renewed interest in exploring their properties.   Here, we investigate statistical properties of the quantum trajectories resulting from heterodyne measurement of fluorescence.  We first present a simple derivation of the measurement result and state backaction of a short heterodyne measurement, and generalize the results to a continuous quantum measurement of the two heterodyne signals.
We proceed to predict the most likely path dynamics of the fluorescing artificial atom between two boundary conditions on the quantum state, focusing on the counterintuitive paths in which energy is {\it gained} during the quantum measurement, as described by Bolund and M{\o}lmer \cite{bolund2014stochastic}.  (Such energy dynamics is also of interest for quantum stochastic thermodynamics \cite{qt1,qt2}.)  We present two independent derivations of this most likely path using the stochastic path integral formalism~\cite{chantasri2013action, chantasri2015stochastic} or quantum control theory \cite{rouchon2014models,rouchon2015efficient}. We also formulate a diagrammatic perturbation theory for computing arbitrary correlation functions for the fluorescence action, and explicitly compute all bilinear and quadratic order correlations to leading order.  Numerical simulation of the system permits the comparison of numerical covariance functions and most likely paths with our theoretical results, showing good agreement.

\begin{figure}[t]
\includegraphics[width=8cm]{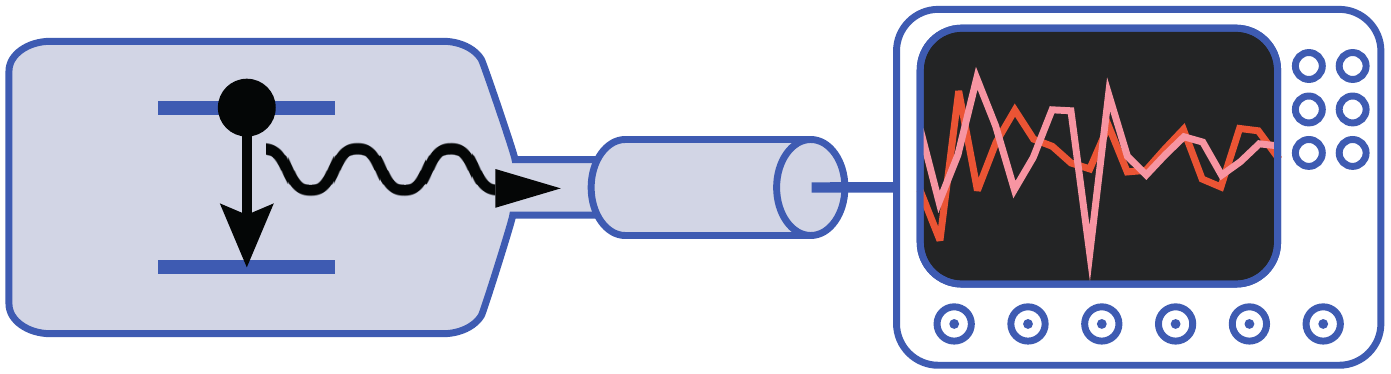}
\caption{A fluorescing artificial atom is indirectly measured by detecting microwave light it produces by spontaneous emission~\cite{Campagne2015}. The fluorescence is preferentially emitted into a single coaxial cable by engineering the electromagnetic environment adequately using a microwave cavity.
By making a heterodyne measurement of the two  electric field quadratures, partial information about the quantum state of the artificial atom is retrieved.}
\end{figure}

The paper is organized as follows.  In section II, we describe the physics of fluorescence, and how to understand the basic physics of measurement.  We then generalize this to a continuous stream of quadrature measurements in time, and derive the equations of motion of the qubit state, given a realization of the continuous measurement.  In Sec.~III, the formalism of the stochastic path integral is introduced, and we derive the stochastic action, stochastic Hamiltonian, and most likely paths of the fluorescence measurement.   For an ideal measurement, we present an analytic solution to the most likely paths in Sec.~IIIA, revealing the anatomy of the fluorescence process.  In Sec.~IV, we reformulate the stochastic differential equations in the It\^o formulation, and derive the associated stochastic action.  From this action, we develop a set of diagrammatic rules for computing arbitrary averages and correlation functions (or covariance functions) in Sec.~V.  We explicitly compute time correlation functions between all stochastic variables, and give approximations to all such correlators of quadratic and bilinear order to up to leading order in the diagrammatic expansion. We compare our results to numerical simulation.  We conclude in Sec.~VI.  Appendix A explores what observables can be constructed with weak fluorescence measurements, and Appendix B generalizes the analysis presented to arbitrary diffusive quantum systems monitored by an arbitrary number of outcomes.

\section{Weak quadrature measurement of fluorescence}

The backaction on the state of a qubit that originates from the detection of its fluorescent emission can be understood by a simple model. Let us define $|g\ra$ and $|e\ra$ as the ground and excited state of the qubit, and $| 0 \ra$ and $|1\ra$ as either the vacuum or single photon excitation of the output field mode that carries the fluorescence signal from the qubit into a waveguide.
We work in the rotating frame (interaction picture) with frequency $\omega_q = (E_e - E_g)/\hbar$, the qubit frequency.
 Owing to the coupling between qubit and output mode, which is characterized by the relaxation rate $\gamma_1$, the two systems get entangled after a time $\delta t$, small compared to $\gamma_1^{-1}$. Starting in an arbitrary state of the form $(a |e\ra + b|g\ra)\otimes |0\ra$, the state changes to
\be
|\psi_f\ra=a \sqrt{1-\epsilon} |e\ra\otimes |0\ra + a \sqrt{\epsilon} |g\ra\otimes |1\ra + b |g\ra\otimes |0\ra.
\label{newstate}
\ee
Here, $\epsilon = \gamma_1 \delta t$ is the probability of a relaxation event when the qubit is excited.
Phase preserving amplification enables the measurement of both quadratures of the output mode (the in-phase and in-quadrature responses at the transition frequency), which acts as a {\it meter} (M) in the measurement of the {\it system} (S) -- the qubit -- owing to its entanglement with the latter. In contrast with photon counting, the measurement of the quadratures of the mode cannot indicate in a single shot whether the output mode contains a photon or not due to quantum zero point fluctuations. Therefore, the measurement of the two mode quadratures leads to two outcomes that are the real and imaginary parts of a random complex field amplitude $\alpha$. This random outcome $\alpha$ both yields some information about the state of the qubit, as well as induces measurement back-action on the qubit. This measurement backaction can be interpreted as physical or informational (Bayesian update) depending on ones point of view of the quantum state.

The probability of the measurement result depends on the state of the qubit. To generalize the treatment to include mixed states we introduce the density matrix of the qubit as
\be
\rho = \begin{pmatrix} \rho_{ee} & \rho_{eg} \\ \rho_{ge} & \rho_{gg}
\end{pmatrix}
= \frac{1}{2} \begin{pmatrix} u & x-iy \\ x+iy & 2-u
\end{pmatrix},
\ee
where we define Bloch sphere coordinates, $x,y$ as usual, but use the excitation $u = 1+z = 2 \rho_{ee}$ as a modified variable that is naturally well suited to the fluorescence measurement problem.  Following amplification and mixing with the heterodyne signal,
the measurement quadrature result $\alpha$ occurs with probability density
\be
P(\alpha) = {\rm Tr}_{S,M}[ \Pi_\alpha |\psi_f\ra \la \psi_f |  \Pi_\alpha],
\ee
where $\Pi_\alpha = |\alpha\ra\la \alpha |$ is a projection operator on coherent state $|\alpha\ra$ and the trace is taken over the system and meter \cite{wiseman2009quantum,Carmichael1993}. Note that $P$ is here the Husimi $Q$ function of the state ${\rm Tr}_{S}[|\psi_f\ra \la \psi_f |]$ \cite{caves2012quantum}.
Evaluating the resulting integrals from tracing in the coherent state basis gives
\be\label{eq-probalpha}
P(\alpha) = e^{-|\alpha|^2} \left[ 1 - \frac{\epsilon u }{2} (1 - |\alpha|^2) + \sqrt{\epsilon} (x {\rm Re}\, \alpha - y {\rm Im }\, \alpha) \right],
\ee
which is normalized with the coherent state measure, which includes the overcompleteness factor of $\pi$, $\int d({\rm Re}\, \alpha) d ({\rm Im}\, \alpha) P(\alpha) /\pi =1$.  This result occurs over a small time step $\delta t$ compared to the duration $T$ of the fluorescence experiment. Consequently, we consider a sequence of many quadrature measurements, giving a stochastic stream of data $\{ \alpha_i \}$, where $i = 1, \ldots, N$, and $N = T/\delta t$. The mean number of photons collected from coherent state in the heterodyne measurement is $|\alpha|^2$, so the photon current of a stream of photons is $\sum_i  |\alpha_i|^2/ (N \delta t)$. This motivates us to define {\it quadrature currents}, each formed from averaging the scaled quadrature variables: $Q, I$. That is, we rescale
\be
{\rm Re}\, \alpha = I \sqrt{\delta t/2}, \quad {\rm Im} \, \alpha = -Q \sqrt{\delta t/2}.  \label{rescale}
\ee
With this transformation,  to first order in $\delta t$, the log-distribution of quadrature currents, $\ln P(I, Q)$ takes this form,
\be
\ln P = -\frac{\delta t}{2} \left(  I^2 - 2 \sqrt{\frac{\gamma_1}{2}} x I + Q^2 - 2 \sqrt{\frac{\gamma_1}{2}} y Q +u \gamma_1\right). \label{logp}
\ee
We see that the distribution of these variables is approximately Gaussian for a small time interval, with mean $\la I \ra= \sqrt{\gamma_1/2}\, x$, and $\la Q \ra = \sqrt{\gamma_1/2}\, y$, and variances of both $Q$ and $I$ given by $1/\delta t$ (with no cross-correlation). This ensures that by averaging these random variables over $N$ realizations, the resulting variance will be reduced to $1/(N \delta t) = 1/T$ (starting from the same initial state). We assume that each measurement within each time interval is statistically independent, consistent with the Markov approximation. Finally, we stress that while it is tempting to view this measurement outcome as a joint weak measurement of two operators $\sigma_x$ and $\sigma_y$ \cite{jordan2005continuous}, (since the real and imaginary part of $\alpha$ are proportional to them on average), this should in fact be viewed as an indirect signature of $\sigma_- = |g\ra \la e| = (\sigma_x - i \sigma_y)/2$, the system operator that drives the interaction (\ref{newstate}).  We discuss this point more in Appendix A, and show that by choosing appropriate contextual values \cite{cv1,cv2}, any system operator may be targeted for repeated weak measurements on the same initial state.

We now focus on the qubit measurement disturbance for a single time step $\delta t$. This projection of the meter state via heterodyne measurement implements a POVM, or generalized measurement, on the qubit \cite{davies1976quantum,kraus1983states,holevo2001statistical,nielsen2010quantum}. Since the strength $\epsilon$ of the measurement is small, this can be classified as a weak measurement. The modified state of the qubit $\rho'_\alpha$, {\it conditioned} on the measurement result $\alpha$ is given by
\be
\rho'_\alpha = {\rm Tr}_M [\Pi_\alpha |\psi_f\ra \la \psi_f |  \Pi_\alpha] / P(\alpha).
\ee
This may also be calculated with the help of the measurement (Krauss) operator ${\cal M}_\alpha$ \cite{kraus1983states}, indexed by the measurement result $\alpha$, and expressed in the $|e\ra, |g\ra$ basis:
\be
{\cal M}_\alpha = \begin{pmatrix} \sqrt{1-\epsilon} & 0  \\ \sqrt{\epsilon} \alpha^\ast & 1 \end{pmatrix} e^{-|\alpha|^2/2}.
\label{mo}
\ee
We can check this implements a POVM on the qubit because $\int (d^2\alpha/\pi)  {\cal M}_\alpha^\dagger {\cal M}_\alpha = {\bf 1}_S$ \cite{wiseman2009quantum,nielsen2010quantum,Carmichael1993,jordan2006qubit}.  The (conditional) state disturbance is therefore given by
$\rho'_\alpha = {\cal M}_\alpha \rho {\cal M}_\alpha^\dagger / {\rm Tr}_M [{\cal M}_\alpha^\dagger {\cal M}_\alpha \rho]$.
Making the required calculations and expressing the results in the modified Bloch coordinates $u, x, y$, we have the following update equations,
\begin{subequations}\label{eq-stateupdate}
\begin{eqnarray}
u'_\alpha &=& u (1-\epsilon)e^{-|\alpha|^2} /P(\alpha), \label{ustep} \\
x'_\alpha &=& \sqrt{1-\epsilon} ( x + \sqrt{\epsilon} u {\rm Re}\,\alpha)e^{-|\alpha|^2}/P(\alpha), \label{xstep}\\
y'_\alpha &=& \sqrt{1-\epsilon} ( y - \sqrt{\epsilon} u {\rm Im}\, \alpha) e^{-|\alpha|^2}/P(\alpha).\label{ystep}
\end{eqnarray}
\end{subequations}
These results are quite interesting, and give great insight into the measurement physics. Note that we have not yet made an expansion in $\epsilon$. We see that if $\epsilon =0$, there is no change of the state.
We see from Eq.~(\ref{ustep}) that without the (re-)normalization factor $P(\alpha)$, the population of the qubit is simply reduced, as is expected from fluorescence, however, the coherences, $x,y$ change stochastically, depending on the results of the measurement. Further, it is clear that counterintuitive state disturbance can occur by looking at the conditions when the fluorescing qubit can {\it increase} in energy \cite{Carmichael1993,bolund2014stochastic}. This happens when $  (u /2) (1 - |\alpha|^2) - (1/\sqrt{\epsilon}) (x {\rm Re}\, \alpha - y {\rm Im }\, \alpha) >1$, which can occur for certain values of $\alpha, x, y, u$.

The physical interpretation of this effect is that if the measurement gives a certain result, the information revealed about the qubit indicates that our prior density matrix now underestimates the current expectation for the population of the qubit, and reassigns it to be at a higher energy. Thus the relaxing qubit can actually increase in energy because of the backaction of the quantum measurement.

It is interesting to note that had we made a direct measurement of the fluorescence with a single photon detector, the energy increase of the system could never happen. After the fluorescence gives state (\ref{newstate}) after time $\delta t$, a photon counting measurement would either obtain a system state collapse to $|g\ra$ if a single photon were detected, or a partial collapse to
\be
\psi' =\frac{1}{\sqrt{|a|^2 (1-\epsilon) + |b|^2}} \begin{pmatrix} a \sqrt{1-\epsilon} \\ b \end{pmatrix},
\label{distcount}
\ee
if a photon is not detected, where we have written the updated state in the $|e\ra, |g\ra$ basis.
This transformation must always {\it decrease} the population of the excited state for any non-zero value of $\epsilon$ regardless of the initial system state $a|e\ra + b|g\ra$.  The partial-collapse of the
qubit state (\ref{distcount}) by indirect measurement also occurs in state-selective tunneling of superconducting phase qubits
\cite{katz2008reversal,katz2006coherent,korotkov2006undoing}.  We see, therefore, that depending on the kind of measurement being done on the fluorescence signal, {\it under exactly the same physical conditions of the qubit}, the backaction on the qubit state is qualitatively different with respect to the qubit energy.

We can turn the update equations \eqref{eq-stateupdate} into a time-local differential equation by scaling the measurement result $\alpha$ as described in Eq.~(\ref{rescale}) and making an expansion of the update equations \eqref{eq-stateupdate} to first order in $\delta t$, to find
\begin{subequations}
\begin{eqnarray}
{\dot u} &=& - \gamma_1 u \left(1 - \frac{u}{2}\right)
- \sqrt{\frac{\gamma_1}{2}} u (x I + y Q), \label{udot}\\
{\dot x} &=& -\frac{\gamma_1}{2} x (1 - u)
+\sqrt{\frac{\gamma_1}{2}} [u I - x ( x I + y Q)], \label{xdot} \\
{\dot y} &=& -\frac{\gamma_1}{2} y (1 - u)
+\sqrt{\frac{\gamma_1}{2}} [u Q - y ( x I + y Q)] \label{ydot}.
\end{eqnarray}
\end{subequations}
These equations can be used to directly find the quantum trajectories of a single run of the experiment, $u(t), x(t), y(t)$ based on the measurement records $I(t)$ and $Q(t)$ of that measurement run. This describes our best estimate of the quantum state given the acquired data. A similar procedure was done in the experiments of Refs.~\cite{murch2013observing,weber2014mapping,Campagne2015}.  These equations of motion are similar in spirit to those of a continuous $Z$ measurement of a qubit \cite{korotkov1999continuous,korotkov2001selective}.
We point out the possibility here also of an increase of $u$ by reversing the sign of the second term in Eq.~(\ref{udot}) so it exceeds the first term with certain $Q$ or $I$ results, raising the energy of the qubit.

For experiments, it is important to generalize these equations to include the effect of pure dephasing, which occurs to the coherences with rate $\gamma_\phi$, as well as the fact that the measurement is not perfectly efficient. We define the efficiency $\eta$ as the ratio of signal collected to the total signal, which reflects the fact that not all of the light is emitted into the coupled waveguide, as well as other sources of loss in the experiment.  The effect of photon loss is accounted for by averaging the system over loss/no loss events, represented by the
operators ${\bf O}_{nl} = \sqrt{\eta} |1\ra\la 1| + |0\ra\la 0|$ in the no-loss case, and ${\bf O}_l = \sqrt{1-\eta}|1\ra\la 1|$ in the loss case.  Averaging state (\ref{newstate}) over these loss events introduces decoherence in the system state.  Since the photon loss operators obey ${\bf O}_{nl}^\dagger {\bf O}_{nl} + {\bf O}_l^\dagger {\bf O}_l = {\bf 1}$, they are a POVM on the light mode \cite{nielsen2010quantum}.  The resulting averaged state is ${\bf O}_{nl} \rho {\bf O}_{nl} + {\bf O}_l \rho {\bf O}_l$, where $\rho = |\psi_f\ra \la \psi_f|$.
Naturally, the meter distribution is only sampled from the no-loss instances, and reduces the (no-loss) norm of the distribution $P(\alpha)$ by an amount $1 - (u/2) \epsilon ( 1- \eta)$.
The effect of extra decoherence is accounted for by applying the phase-flip superoperator of strength $\gamma_\phi \delta t$ to the qubit state.  Revisiting the derivations for the distribution of the meter results $\alpha$ and the state disturbance in the small $\delta t$ limit, gives a Gaussian distribution of mean $\la I \ra = \sqrt{\gamma_1 \eta/2} x$, $\la Q \ra = \sqrt{\eta \gamma_1/2} y$, and variances ${\rm Var} [I] = {\rm Var} [Q] = 1/\delta t$.  These considerations generalize the quantum trajectory equations to:
\begin{subequations}\label{eq-mlpwitheta}
\begin{eqnarray}
\hspace*{-2.em}{\dot u} &=& - \gamma_1 u \left(1 - \eta \tfrac{u}{2}\right)
- \sqrt{\tfrac{\eta \gamma_1}{2}} u (x I + y Q), \label{udotg} \\
\hspace*{-2.em}{\dot x} &=& -\tfrac{\gamma_1}{2} x (1 - \eta u) - \gamma_{\phi} x
+\sqrt{\tfrac{\eta \gamma_1}{2}} [uI-x(x I +y Q)], \label{xdotg}  \\
\hspace*{-2.em}{\dot y} &=& -\tfrac{\gamma_1}{2} y (1 - \eta u) - \gamma_{\phi} y
+\sqrt{\tfrac{\eta \gamma_1}{2}} [uQ-y(xI+yQ)]. \label{ydotg}
\end{eqnarray}
\end{subequations}

\section{Stochastic action and most likely path}
The equations of motion (\ref{eq-mlpwitheta}) for the quantum state are in their most physical form - the usual rules of calculus apply, and there is a clear physical interpretation of the results.
We can make them explicitly stochastic differential equations by choosing the values $I, Q$ at random from the distribution (\ref{logp}). These stochastic differential equations must be interpreted in a Stratonovich form. Rather than pursue this route, we introduce a functional approach to the dynamics by considering the time-evolution operator of densities of these stochastic trajectories with a stochastic path integral, following the strategy outlined in Refs.~\cite{chantasri2013action,chantasri2015stochastic}. We implement the equations of motion (\ref{eq-mlpwitheta}) at every time interval with associated Lagrange multipliers $p_u, p_x, p_y$ which take on their own dynamics. These $p$ variables may also be interpreted as canonically conjugate to the stochastic variables $u, x, y$.  Any functional of trajectories $u(t), x(t), y(t)$, or the measurement results $I(t), Q(t)$, can be averaged over the ensemble with functional integrals over all variables $u, x, y, p_u, p_x, p_y, Q, I$ of the stochastic path integral defined by a stochastic action given by
\begin{subequations}\label{eq-actionandH}
\begin{eqnarray}
{\cal S} &=& \int dt'\{ -p_u{\dot u} - p_x {\dot x} - p_y {\dot y} + {\cal H}\}, \label{S} \\
{\cal H} &=&  -p_u [\gamma_1 u (1- \eta u/2) + \zeta u (x I +y Q)]  \label{H} \nonumber \\
&-& p_x[(\gamma_1/2) x(1-\eta u) + \gamma_\phi x + \zeta (-u I + x(xI +yQ))] \nonumber  \\
&-& p_y[ (\gamma_1/2) y (1-\eta u) + \gamma_\phi y  + \zeta(- u Q + y (x I + y Q))] \nonumber \\
&-& I^2/2 + \zeta x I - Q^2/2 + \zeta y Q - \eta \gamma_1 u/2.
\end{eqnarray}
\end{subequations}
Here we have simplified notation by introducing $\zeta =  \sqrt{\eta \gamma_1/2}$.  The first three lines of (\ref{H}) capture the equations of motion (\ref{eq-mlpwitheta}), while the last line of (\ref{H}) is the log-probability of $I, Q$ in the presence of finite efficiency $\eta$.  We may also introduce boundary conditions (or other constraints) on the dynamics by introducing other terms to the action of the form, $B = \int dt' r_{i,0} p_i(t') \delta(t')$, where $r_i= (u, x, y)$ (with $i = u,x,y$) for the initial condition, and a similar such term for the final condition. The action has a canonical form, with a \textit{stochastic Hamiltonian} $\cal H$.  In Appendix B, we generalize this analysis to account for an arbitrary diffusive quantum system that is continuously monitored.

\begin{figure}
\includegraphics[width=8.5cm]{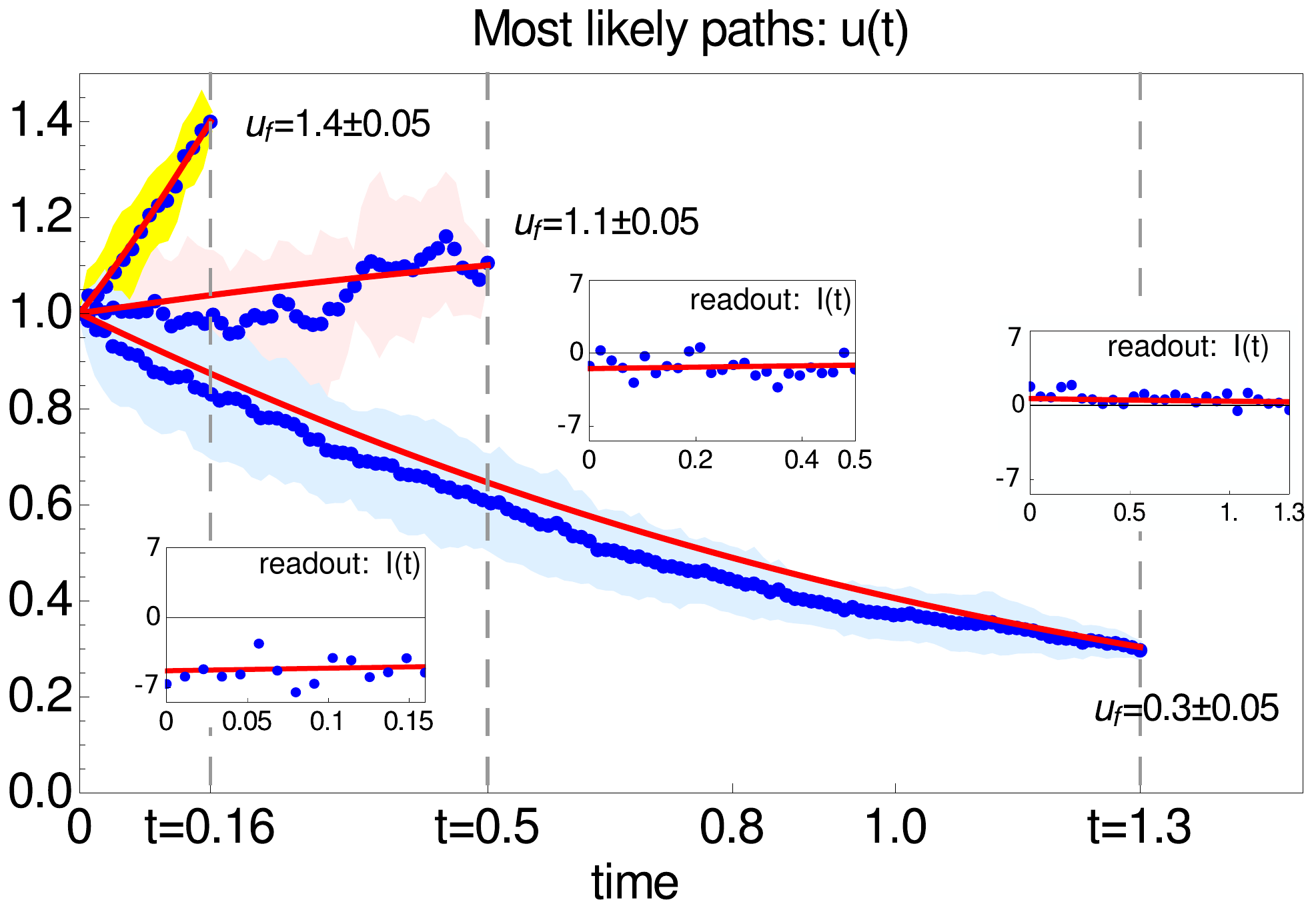}
\caption{The most likely paths predicted from the stochastic action principle (solid red curves) are shown in comparison with the most probable paths derived from simulated $10^5$ trajectories, all have their initial states at the ${\hat x}$-state: $(u_i, x_i, y_i) = (1, 1, 0)$. We consider subensembles of trajectories with fixed final states, where three sets of final states are: $u_f(t_f = 0.16) = 1.4$, $u_f(t_f = 0.5) = 1.1$, and $u_f(t_f = 1.3) = 0.3$ (time is given in units of $\gamma_1^{-1}$), with a tolerance of size $0.05$. The most probable paths are extracted from the trajectory data by using the minimal average trace distance (see text). The (colored) shaded regions shows one standard deviation band as a function of time from the simulated most probable paths. Here we only show the paths in the $u$-variable as well as the $I$-quadrature readout, plotted in units of $\sqrt{\gamma_1/2}$, as insets. The most likely evolution in the $y$ and $Q$-quadrature are given by $y = Q =0$ for the chosen boundary conditions, and the evolution in $x$ can be inferred from $u$ because the qubit states stay pure in this particular case: $\eta = 1$, $\gamma_{\phi} = 0$. }
\label{fig-MLPsimulation}
\end{figure}

We now consider the most likely path between two boundary conditions on the state, one at the beginning of the measurement, and another at the end of the measurement.
This most likely path will be the one that captures the largest fraction of trajectories in some small neighborhood of all the trajectories that obey the imposed boundary conditions. The theory of the most likely path for quantum trajectories was experimentally verified for a continuous $Z$ measurement with Rabi drive applied in the experiment of Ref.~\cite{weber2014mapping}. The most likely path is technically found by maximizing the global probability of all paths between the boundary conditions, which may be cast as a stochastic action principle \cite{chantasri2013action}.
Taking functional derivatives to extremize the action, $\delta S=0$, yields the equations of motion and constraints
for the most likely path.

  The $p_u, p_x, p_y$ functional derivatives recover the original equations \eqref{eq-mlpwitheta}, but with the interpretation of all the variables as the most likely ones, rather than as stochastic variables.  The $u, x, y$ functional derivatives gives the equations of motion for the variables $p_u, p_x, p_y$,
\begin{subequations}\label{eq-mlp2witheta}
\begin{eqnarray}
{\dot p}_u &=& \gamma_1 p_u (1 - \eta u) - \gamma_1 \eta (p_x x + p_y y)/2 \nonumber\\
&+& \zeta [p_u(x I + yQ) - p_x I - p_y Q] +\eta \gamma_1/2, \\
{\dot p}_x &=& \gamma_1 p_x (1- \eta u)/2 + \gamma_\phi p_x \nonumber \\
&+&  \zeta (p_u u I + 2 p_x x I + p_x y Q + p_y y I-I),  \\
{\dot p}_y &=& \gamma_1 p_y (1 - \eta u)/2 + \gamma_\phi p_y \nonumber \\
&+& \zeta ( p_u u Q + p_x x Q + p_y x I + 2 p_y y Q-Q),
\end{eqnarray}
\end{subequations}
In the case of phase preserving measurement, we have two quadrature variables $I, Q$, and consequently have two constraints, linking the most-likely path variables through the most-likely values of the quadrature variables via functional $I,Q$ derivatives, $\delta {\cal S}/\delta I, \delta {\cal S}/\delta Q=0$,
\begin{subequations}
\begin{eqnarray}
I/\zeta &=& x + p_x (u -x^2) - p_u u x - p_y x y, \\
Q/\zeta &=& y + p_y (u - y^2) - p_u u y - p_x x y.
\end{eqnarray}
\end{subequations}

The six equations of motion for the most likely path involving $u, x, y, p_u, p_x, p_y$ in equations~\eqref{eq-mlpwitheta} and \eqref{eq-mlp2witheta}, combined with the two constraints for $I, Q$ may be solved applying boundary conditions for the initial and final values of $u_0, x_0, y_0$ and $u_f, x_f, y_f$, separated by a given time $t$.  This is possible because the six first order differential equations admit six constants of integration. 

In Figure~\ref{fig-MLPsimulation}, we plot the predicted most likely paths in comparison with numerically simulated qubit trajectories, where the most likely paths in the latter case are obtained by selecting subensemble of trajectories with chosen final states and then searching for paths with minimum average trace distances between any two trajectories (for more detail, see Ref.~\cite{weber2014mapping}) to calculate the most probable paths. We find good agreement both for the qubit state variables and for the most likely measurement readouts.  The most likely path equations are generalized to an arbitrary diffusive, continuously monitored system in Appendix B.

\begin{figure}
\includegraphics[width=8.5cm]{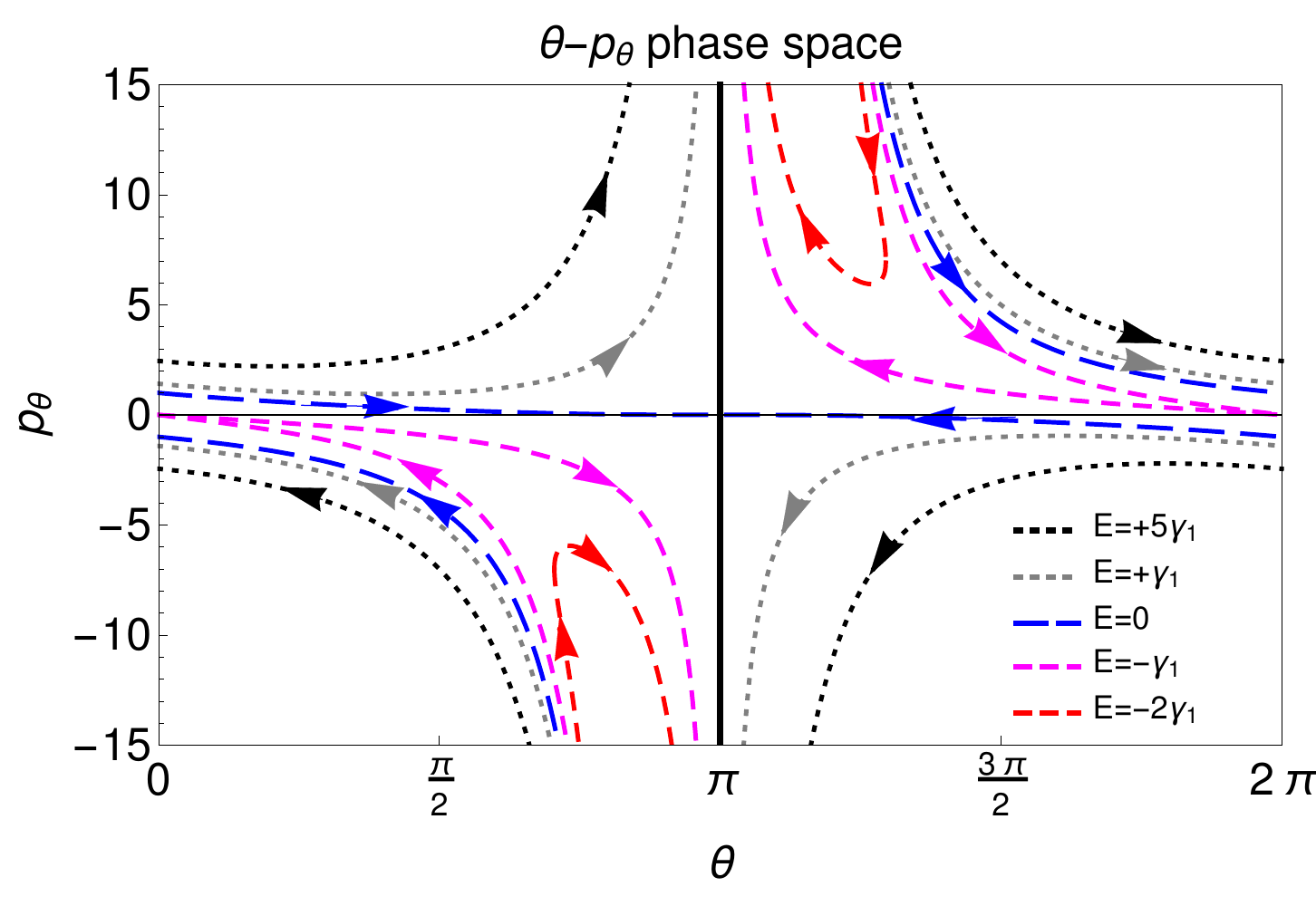}
\caption{Anatomy of fluorescence:  This phase space diagram represents all possible most-likely paths between two boundary conditions in the ideal measurement case.  The horizontal axis represents the polar angle $\theta$ of pure states on the $x-z$ great circle of the Bloch sphere, where the angle is measured from the $z$ axis.  Thus $\theta = 0, 2\pi$ represents the excited state $|e\ra$, while $\theta = \pi$ represents the ground state $|g\ra$.  The vertical axis is the associated canonically conjugate variable, $p_\theta$.
The most likely paths are fixed by the stochastic energy $E$, and plotted for a choice of several energies.  The dashed blue paths are the most likely given fixed endpoints, which correspond to the zero energy lines. }
\label{fig-phasespace}
\end{figure}

\subsection{The anatomy of fluorescence: Analytic solution for the ideal measurement}
In some cases, the most likely path equations may be solved exactly. In any case, there is a constant of motion $E = {\cal H}$, fixed as the stochastic energy of the dynamics once the boundary conditions and the time between them is fixed. There may be other constants of motion.

We now consider a special case, by noticing that a particular solution $y=0$ for all time will satisfy the most likely path equations for $y$ and $p_y$, provided that the most likely value $p_y=0$ for all time, which then implies that the most likely value of $Q=0$ as well.  We stress that this is not equivalent to integrating over $Q$.  That procedure discards information, and produces different equations of motion for the state that leads to decoherence even if there is no extra dephasing.  Making the simplification of the most likely values $y=p_y = Q =0$ allows us to recast the remaining equations of motion as one dimensional under certain assumptions.
We consider an ideal measurement set-up, where $\gamma_\phi = 0$, and $\eta=1$.  In this case, the quantum measurement dynamics remains pure the whole time.  This gives an additional constant of motion, because $x^2 + z^2 = x^2 + (1-u)^2 = 1$, and we may therefore replace $x = \sin \theta$, and $u = 1 + \cos \theta$.  The polar angle $\theta$ from the $z$ axis completely characterizes the dynamics of the quantum state in this ideal situation.  Making this transformation to either of the stochastic differential equations (\ref{udot}) and (\ref{xdot}) gives a single equation to solve,
\be
{\dot \theta} = \frac{\gamma_1}{2} \sin \theta + \sqrt{\frac{\gamma_1}{2}} (1 + \cos \theta) I. \label{thetaeq}
\ee
For this one degree of freedom, the simplified stochastic action and stochastic Hamiltonian is
\begin{subequations}
\begin{eqnarray}
{\cal S}_\theta &=& \int dt'\{ -p_\theta {\dot \theta} + {\cal H}_\theta \}, \label{Stheta} \\
{\cal H}_\theta &=&  \frac{\gamma_1}{2} p_\theta \sin \theta
-\frac{\gamma_1}{2}(1+\cos \theta)
\nonumber \\ &+&  \sqrt{\frac{\gamma_1}{2}} p_\theta (1+\cos \theta) I  - I^2/2 +\sqrt{\frac{\gamma_1}{2}} I \sin \theta.  \label{Htheta}
\end{eqnarray}
\end{subequations}
Here, we have introduced the canonically conjugate variable $p_\theta$ as a Lagrange multiplier variable to implement the state update constraint.

In this special case of one dimension, we can solve the equations of motion for the most likely path in closed form.  Let us first find the most likely quadrature current from $\delta S/\delta I =0$, giving
\be\label{eq-mlpreadout}
I = \sqrt{\frac{\gamma_1}{2}}(\sin \theta + p_\theta (1+\cos \theta)).
\ee
Since we can always reconstruct the time-dependence of $I$ from that of $\theta(t), p_\theta(t)$ from this result, we insert the most likely $I$ back into the action (\ref{Stheta}) to find the modified Hamiltonian of
\begin{eqnarray}
{\cal H}_\theta' &=& \gamma_1 (a p_\theta^2 + b p_\theta + c), \\
a &=& -c = \cos(\theta/2)^4, \quad b = \sin \theta \left( 1 + \frac{\cos \theta}{2}\right). \nonumber
\end{eqnarray}
We can now find the most-likely paths in a simple way, because of the constant stochastic energy, ${\cal H}_\theta' = E$.  The stochastic energy $E$ is fixed by setting the initial condition, the final condition, and the time elapsed between them.  We can solve for the lines of motion $p_{\theta,E}(\theta)$ in the phase space for any energy to give two solutions of the quadratic equation (see Fig.~\ref{fig-phasespace}).  Of particular interest are the \textit{zero energy lines}, corresponding to the most probable dynamics between fixed endpoints,
\be
{p}_{\theta,E=0}^{\pm}(\theta) = \frac{\pm \cos(\frac{\theta}{2})\sqrt{10+6 \cos\theta} - (2 + \cos \theta) \sin \theta}{(1 + \cos \theta)^2}.
\ee
The solution ${p}_{\theta,E=0}^+$ corresponds to the paths that decrease in energy, relaxing the qubit to the ground state $|g\ra$ at $\theta = \pi$, whereas ${ p}_{\theta,E=0}^-$ describe paths that increase in energy, exciting the qubit state to the state $|e\ra$ at $\theta= 0$.
At $\theta=0$, the solution limits to ${p}_{\theta,E=0}^\pm = \pm 1$, whereas when $\theta \rightarrow \pi$, ${p}_{\theta,E=0}^+$ limits to $0$ but the minus solution diverges as ${p}_{\theta,E=0}^- \sim 8/(\theta - \pi)^3$.

\begin{figure}
\includegraphics[width=8.1cm]{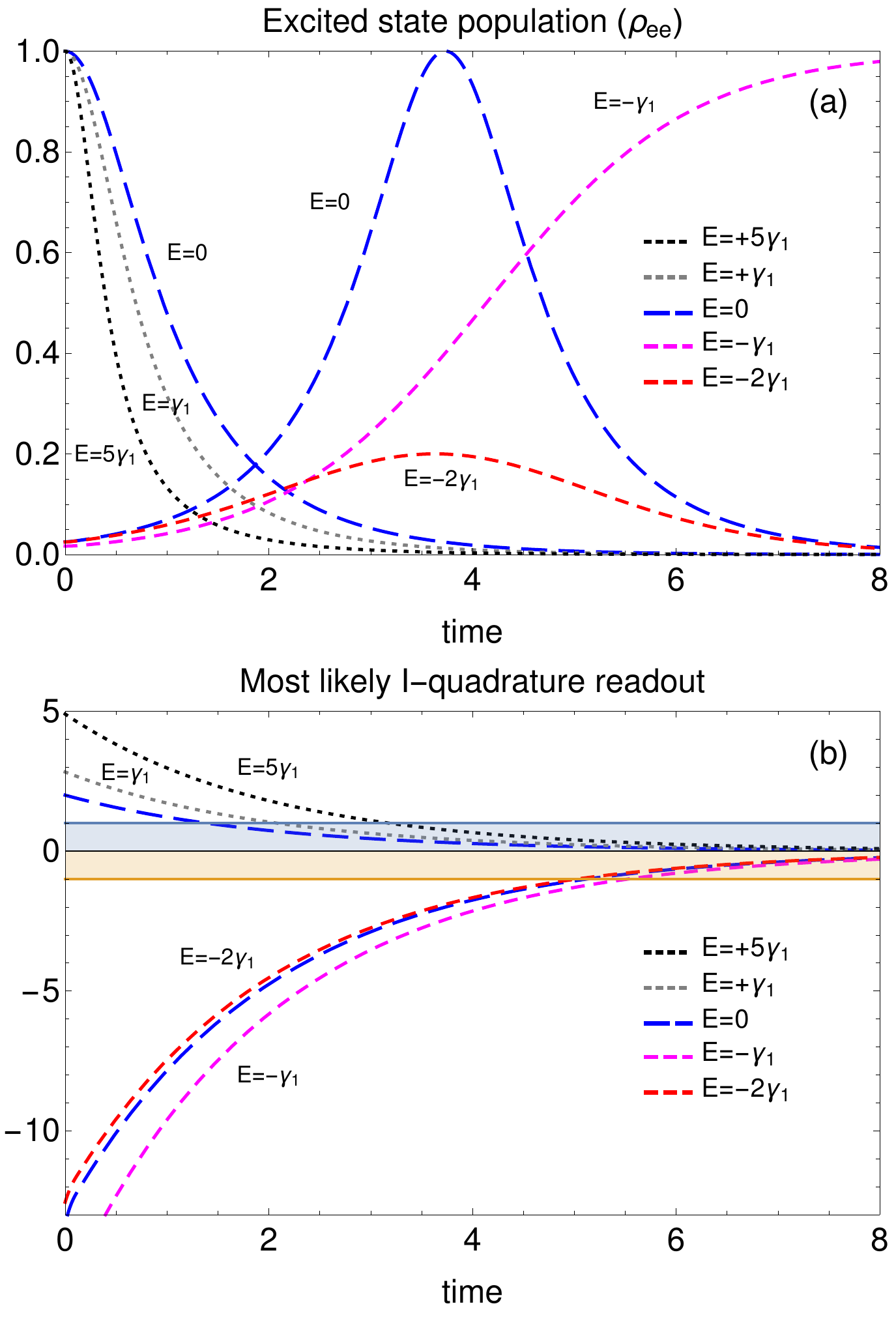}
\caption{The most likely path solutions for different values of stochastic energies chosen in Figure~\ref{fig-phasespace}. Panel (a) shows the excited state population $\rho_{ee}$ as functions of time, from $t_0 =0$ to a final time $t_f = 8$ (time is shown in units of $\gamma_1^{-1}$). There are three decaying solutions shown in dashed blue (the left one), dotted grey, and dotted black curves, losing energy from the excited state $\theta_0 = 0$ to the ground state $\theta_f = \pi$. The other zero-energy solution (blue dashed curve in the middle) is obtained with the initial and final states $\theta_0 = 0.9 \pi, \theta_f = -0.9 \pi$, showing a temporary excitation followed by a decay back to the ground state. The magenta curve has its initial state at $\theta_0 = 0.92 \pi$ and its final state at $\theta_f = 0.08\pi$, showing that it asymptotically reaches the excited state. The red curve is obtained from $\theta_0 = 0.9 \pi$ to $\theta_f = \pi$. Panel (b) shows the corresponding most likely $I$-quadrature readouts as functions of time, plotted in units of $\sqrt{\gamma_1/2}$, for the same set of stochastic energies and boundary states. The orange and blue shaded regions correspond to the regions of $-1\le x \le 0$ and $0 \le x \le 1$, respectively.  Values of $I$ outside of that range are a continuous analog of the weak value effect \cite{aharonov1988result}.}
\label{fig-mlpevolution}
\end{figure}

The advantage of having the lines of constant stochastic energy is that the action may be directly computed,
\be
{\cal S}_\theta = \int dt' (-p_\theta {\dot \theta} + {\cal H}_\theta)
 =-  \!\! \int \!\! p_{\theta,E} d\theta + E T,
 \ee
where the first integral is from $t=t_0$ to $t_f$ and the second integral is from $\theta = \theta_0$ to $\theta_f$.
In the case of the zero energy lines, the first term fixes the action entirely.  The action integral has an analytic form of
\begin{eqnarray}
{\cal S}_{\theta,E=0} &=& -\!\! \int \!\!p_{\theta,E=0}^\pm d\theta =
\frac{\mp \sqrt{10+6 \cos \theta} \sin(\theta/2) +2}{2(1+\cos\theta)} \nonumber \\
&-& 2 \ln \cos(\theta/2) \mp 2 \tanh^{-1}\frac{\sqrt{2}\sin(\theta/2)}{\sqrt{5 + 3\cos \theta}}.
\end{eqnarray}
When evaluated between the initial and final condition, $\theta_0$ and $\theta_f$, the action ${\cal S}_{\theta, E=0}$ approximates the log-probability of this path.
The elapsed time $T$ of the most likely path may be explicitly given as a function of the stochastic energy. This is given in the case of the zero energy solutions by
\begin{eqnarray}
T_{E=0} &=& \int d\theta/|{\dot \theta}_{E=0}| = \int d\theta |\partial_{p_\theta} {\cal H}_\theta|^{-1} \nonumber \\
&=& \int d\theta \frac{\sqrt{2}}{\gamma_1 \cos (\theta/2)\sqrt{5+3\cos \theta}} \nonumber \\
&=& \frac{2}{\gamma_1} \tanh^{-1}\left(\frac{\sqrt{2}\sin (\theta/2)}{\sqrt{5 - 3\cos \theta}}\right),
\end{eqnarray}
which is evaluated between ${\theta_0}$ and ${\theta_f}$ for $E=0$.  This later equation can be inverted to find the time dependence of the zero-energy line ${\theta_{E=0}}(T)$, which is then used in computing the most likely evolution $\theta_{E=0}(t)$ for $t = t_0$ to any final time $t_0 +T$.

The most likely paths computed with the procedure presented above are plotted in Fig.~\ref{fig-phasespace} for several different choices of stochastic energy. The two zero energy solutions are in blue dashed lines, describing the most likely ways to either lose or gain energy in the measurement process.
In panel (a) of Fig.~3, we plot the most likely path of $\rho_{ee}$ versus time, and in panel (b) we show the most likely quadrature readout $I$ versus time in Eq.~\eqref{eq-mlpreadout}, both for the same choices of stochastic energies.

Also of interest is the solution of (the most likely angle) $\theta = 0$ for all time, corresponding to stabilizing the excited state $|e\ra$.
In this case, Eq.~(\ref{thetaeq}) is satisfied if $I=0$ for the most likely $I$ quadrature value.  The solution has an action given entirely by the stochastic energy,
$S_\theta = E t = - \gamma_1 t$, so the probability of this solution decays exponentially in time with rate $\gamma_1$.  This behavior is as one might have expected from Eq.~(\ref{newstate}).

\begin{table*}
{\renewcommand{\arraystretch}{1.8}
\begin{tabular}{  |>{\centering\arraybackslash} m{2cm} |>{\centering\arraybackslash} m{3.6cm} | >{\centering\arraybackslash} m{4.4cm}|  >{\centering\arraybackslash} m{2.5cm} | }
\hline
Type & Labels of vertices & Full forms & Diagrams  \\ \hline
Initial & $ x_0, y_0, u_0$ & $ r_{i,0}\! \int \!  dt'\, p_i(t') \delta(t') $ & \begin{tikzpicture}[node distance=0.6cm and 0.8cm]
\coordinate (b2);
\coordinate[right=0.5cm of b2] (bp);
\draw[particle] (b2) -- (bp);
\draw (bp) circle (.06cm);
\end{tikzpicture} \\ \hline
3 legged  & $p_{x} u\, \xi_I, p_{y} u \xi_Q $ & $\zeta \! \int \!dt' p_i(t') u(t')\, \xi_k(t') $ & \begin{tikzpicture}[node distance=0.3cm and 0.5cm]
\coordinate (b2);
\coordinate[right=0.5cm of b2] (bp);
\coordinate[below=0.35cm of bp](bpp);
\coordinate[right=of bp](c0);
\draw[particle] (b2) -- (bp);
\draw[gluon](bp)--(bpp);
\draw (bp) circle (.06cm);
\draw[particle2](bp) sin (c0);
\end{tikzpicture} \\ \hline
4 legged & $p_u u x \xi_I$, $p_u u y \xi_Q$, $p_x x^2 \xi_I $, $p_x x y \xi_Q$, $p_y y x \xi_I$, $p_y y^2 \xi_Q$& $-\zeta \!\int \! dt' \, p_i(t') r_i(t') r_j(t') \xi_k(t')$ &
\begin{tikzpicture}[node distance=0.3cm and 0.5cm]
\coordinate (b2);
\coordinate[right=0.5cm of b2] (bp);
\coordinate[below=0.5cm of bp](bpp);
\coordinate[right=of bp](c0);
\coordinate[below=0.2cm of c0](c1);
\coordinate[above=0.2cm of c0](c2);
\draw[particle] (b2) -- (bp);
\draw[gluon](bp)--(bpp);
\draw (bp) circle (.06cm);
\draw[particle2](bp) sin (c1);
\draw[particle2](bp) sin (c2);
\end{tikzpicture}  \\ \hline
\end{tabular}}
\label{table1}
\caption{Different possible vertices and associated diagrams: initial vertices, three-legged, and four legged vertices. The index $i$(or $j$) represents the system variables: $r_i = u, x, y$ for $i=u,x,y$, while the index $k$ represents the noise variables: $k= I, Q$. In the second column, we list all possible types of vertices, corresponding to all terms in the interaction action ${\cal S}_I$ in Eq.~\eqref{eq-interaction}.}
\end{table*}

\section{It\^{o} formulation and stochastic action}
With the stochastic action, in Eqs.~\eqref{eq-actionandH}, we can also consider different kinds of correlation functions, such as $\la u(t_1) u(t_2) \ra$, $\la I(t_1) Q(t_2)\ra$, etc., particularly when we fix the first boundary only (only fix the initial state). For this purpose, it is convenient to transition from the Stratonovich form to an It\^o form of the stochastic differential equations (SDEs). This is equivalent to evaluating the state update from the beginning of the time step rather than the middle of the time step. The transition may be made by writing the quadrature currents as their averages, plus stochastic variables, $I(t) = \zeta x + \xi_I$ and $Q(t) = \zeta y + \xi_Q$, where $\xi_I$ and $\xi_Q$ are independent, zero mean, white Gaussian variables, of variance $1/\delta t$. In addition to this, the conversion of Stratonovich differential equations to It\^o differential equations demands an additional drift term in general \cite{gardiner1985handbook}.
Let $r_i$ represent the components of vector $(u, x, y)$ for $i=u,x,y$ and $\xi_j$ represent the components of the vector $(\xi_I, \xi_Q)$ for $j = 1,2$.
Then, a Stratonovich differential equation of the form
\be
{\dot r}_i = f_i + \sum_j L_{ij} \xi_j
\ee
is equivalent to It\^o SDEs of the same type \cite{gardiner1985handbook}, but with an additional drift term, $d_i = (1/2) \sum_{j,k} (\partial_{r_k} L_{ij}) L_{kj}$.
This transformation gives the It\^o equations of motion,
\begin{subequations}\label{eq-itoequations}
\begin{eqnarray}
{\dot u} &=& - \gamma_1 u
- \sqrt{\frac{\eta \gamma_1}{2}} u (x \xi_I + y \xi_Q), \\
{\dot x} &=& -\left(\frac{\gamma_1}{2} + \gamma_{\phi}\right) x
+\sqrt{\frac{\eta \gamma_1}{2}} \left[u \xi_I - x ( x \xi_I + y \xi_Q)\right],  \\
{\dot y} &=& -\left(\frac{\gamma_1}{2} + \gamma_{\phi}\right) y
+\sqrt{\frac{\eta \gamma_1}{2}} \left[u \xi_Q - y ( x \xi_I + y \xi_Q)\right].
\end{eqnarray}
\end{subequations}
which is equivalent to the one written in Ref.~\cite{wiseman2009quantum,Campagne2015} in component form.
The It\^o form of the SDEs is nice since we may directly find the averages of each variable by dropping the stochastic term, and solving the equations for the averages (we can also find these equations in the Stratonovich form by taking $\eta \rightarrow 0$),
\be
\la u(t) \ra = u_0 e^{-\gamma_1 t}, \quad \la x(t) \ra = x_0 e^{-\gamma_2 t},
\quad \la y(t) \ra = y_0 e^{-\gamma_2 t}, \label{aves}
\ee
as expected. Here, $\gamma_2 = \gamma_1/2 + \gamma_\phi$ is the total dephasing rate, combining the effects of relaxation and pure dephasing. To go beyond the averages, it is convenient to introduce the stochastic path integral for this formulation of the physics. Following the prescription of Ref.~\cite{chantasri2015stochastic} we have a new stochastic action $\tilde {\cal S}$ of the form,
\begin{subequations}
\begin{eqnarray}
{\tilde {\cal S}}&=& \int dt'  [ -p_u {\dot u} - p_x {\dot x} - p_y {\dot y} + {\tilde {\cal H}}], \\
{\tilde {\cal H}} &=& - p_u [ \gamma_1 u + \zeta u (x \xi_I + y \xi_Q)]
\nonumber \\
&-& p_x [\gamma_2 x + \zeta (-u \xi_I + x(x \xi_I + y \xi_Q))] \nonumber \\
&-& p_y [\gamma_2 y + \zeta (-u \xi_Q + y ( x \xi_I + y \xi_Q))]  \nonumber \\
&-& \xi_I^2/2 - \xi_Q^2/2.
\end{eqnarray}
\end{subequations}
where $\tilde {\cal H}$ is the new stochastic Hamiltonian, capturing the equations of motion of the state \eqref{eq-itoequations} and the log-probability of noise terms $\xi_I$ and $\xi_Q$.

\section{Diagrammatics}
We can now develop the diagrammatic way to compute covariance functions.  The stochastic action is split into two parts, the first containing terms that are quadratic in the various variables, and the rest containing the higher order terms.  The quadratic order terms are the \textit{free action}, given by
\begin{eqnarray}
\tilde{{\cal S}}_F &=& \int dt' \big\{- p_u ({\dot u} + \gamma_1 u) \nonumber
- p_x ({\dot x} + \gamma_2 x) - p_y ({\dot y} + \gamma_2 y)  \\
\label{eq-freeaction} &&\qquad  - \xi_I^2/2 - \xi_Q^2/2 \big\},
\end{eqnarray}
The rest of the terms are proportional to $\zeta$, and give the \textit{interacting action},
\begin{align}\label{eq-interaction}
\tilde{{\cal S}}_I = B\, -\, &\zeta\!\! \int \! dt' (p_u u + p_x x + p_y y) (x \xi_I + y \xi_Q) \nonumber \\
& + \zeta \!\!\int \! dt' u(p_x \xi_I + p_y \xi_Q),
\end{align}
including the initial boundary terms 
\be
B = \!\int \!\!dt' \{ u_0 p_u  \delta(t') + x_0 p_x \delta(t') + y_0 p_y \delta(t')\}.
\ee
We will then write these actions in terms of propagators and define interacting vertices, in order to use diagrammatic rules to construct solutions for the statistical averages.

\subsection{Propagators}
From the form of the free action in Eqs.~\eqref{eq-freeaction}-\eqref{eq-interaction}, the dynamics is already diagonal in our chosen variables, so we can write the free action in terms of propagators $\tilde{{\cal S}}_F = - \int dt dt' p_i(t)G_i^{-1}(t,t')r_i(t')  - \frac{1}{2}\int dt dt' \xi_j(t) G_{j,k}^{-1}(t,t') \xi_k(t')$, where the inverse Green functions for the state variables satisfy
\be
G_i^{-1}(t, t') = \delta(t-t')\left( \frac{d}{dt} + \gamma_i\right) ,
\ee
for $i= u, x, y$, $\gamma_u = \gamma_1$, and $\gamma_{x, y} = \gamma_2$. The Green functions are consequently given by
\be
G_i(t, t') = \la r_i(t) p_i(t')\ra_F = \Theta(t - t') e^{-\gamma_i (t - t')},
\ee
where the angle bracket $\la \ldots \ra_F$ denotes the two-time covariance function taken with respect to the free action, and the $\Theta(t)$ function is a left continuous Heaviside step function ($\Theta(0)=0$ and $\lim_{t\rightarrow 0^+} \Theta(t) = 1$, see Ref.~\cite{chantasri2015stochastic}).  By definition, the white noise sources are statistically independent and delta-correlated in time, so we have the Green function for the noise term,
\be
G_{j,k}(t_1, t_2) = \la \xi_j(t_1) \xi_k(t_2) \ra_f = \delta_{jk} \delta(t_1-t_2).
\ee
where here, the indices are $j,k= I, Q$ denoting the two noise terms of the two quadratures.

\subsection{Vertices}
We see from the interacting action that vertices come in two basic types, based on the number of legs they have, either 3 or 4 in our case, plus the initial condition (we do not consider a final condition here, but could).  These verticies are given in the Table I.  The diagrammatic rules are given in Ref.~\cite{chantasri2015stochastic}, and are here adapted to these new verticies and propagators.

\subsection{Averages and correlation functions}
We illustrate the diagrammatic rules first with the average of the coordinates,
\begin{eqnarray}
\nonumber \la r_i(t) \ra&=& \la r_i(t) e^{\cal S}\ra_F =
\begin{tikzpicture}[node distance=0.6cm and 0.8cm]
\coordinate[label=above:$r_{i,t}$] (b2);
\coordinate[right=of b2,label=above:$r_{i,0}$] (bp);
\draw[particle] (b2) -- (bp);
\draw (bp) circle (.06cm);
\fill[black] (b2) circle (.05cm);
\end{tikzpicture}  \\
&=& \la r_i(t) r_{i,0} \int dt'  p_i(t') \delta(t') \ra_F \nonumber \\
&=& r_{i,0} \int dt' \Theta(t-t') e^{-\gamma_i (t - t')} \delta(t') \nonumber  \\
&=& r_{i,0} e^{-\gamma_i t},
\end{eqnarray}
for the three choices of system coordinates $i = u,x,y$. This recovers our earlier results (\ref{aves}), the exponential decays, for the averages.

We move on to compute the correlation functions of the form $\la r_i(t_1) r_j(t_2) \ra$ (between system variables), as well as $\la r_i(t_1) \xi_k(t_2) \ra$  (between state variables and noise variables). We note that correlation functions including the detector output variables, quadrature $I$ and $Q$, can be found by writing $I = \zeta x + \xi_I$ and $Q = \zeta y + \xi_Q$, and decomposing the correlation function into ones of the form between the system variables and the noises.

Since the series does not truncate in general, we will calculate the first few tree-level diagrams to approximate the correlation functions. As was shown in Ref.~\cite{chantasri2015stochastic} this approximation represents a small-noise approximation, and works well when the trajectories bunch around the average (the same is the case when the saddle-point approximation to the path integral well approximates its value). In this case, the parameter $\zeta$ which is in front of all the interacting diagrams involves the square-root of the efficiency $\eta$, which provides an additional expansion parameter (in the experiments of Ref.~\cite{Campagne2015} $\eta \approx 0.24$) since each vertex brings a factor of $\zeta$.

We begin with the correlation function of $u(t)$ at times $t_1$ and $t_2$,
\begin{align}
&\la u(t_1) u(t_2)\ra \nonumber\\
&=  \begin{tikzpicture}[node distance=0.6cm and 0.8cm]
\coordinate[label=below:$u_{t_2}$] (b2);
\coordinate[right=of b2,label=below:$u_0$] (bp);
\coordinate[above=of b2,label=above:$u_{t_1}$](a1);
\coordinate[above=of bp,label=above:$u_0$](ap);
\draw[particle] (b2) -- (bp);
\draw[particle] (a1) -- (ap);
\draw (bp) circle (.06cm);
\draw (ap) circle (.06cm);
\fill[black] (b2) circle (.05cm);
\fill[black] (a1) circle (.05cm);
\end{tikzpicture} +
 \begin{tikzpicture}[node distance=0.6cm and 0.8cm]
\coordinate[label=below:$u_{t_2}$] (b2);
\coordinate[right=of b2,label=below:$x$,label=above right:$u$] (bp);
\coordinate[above=of bp,label=below left:$I$,label=above:$x$,label=below right:$u$] (ap);
\coordinate[left=1.1cm of ap,label=above:$u_{t_1}$] (a1);
\coordinate[right=of ap](cc);
\coordinate[right=of bp](bb);
\coordinate[above=0.3cm of cc,label=right:$x_0$](cp);
\coordinate[below=0.1cm of cc,label=right:$u_0$](dp);
\coordinate[above=0.1cm of bb,label=right:$u_0$](cpp);
\coordinate[below=0.3cm of bb,label=right:$x_0$](dpp);
\draw[particle] (a1) -- (ap);
\draw[particle] (b2) -- (bp);
\draw[gluon] (ap) --  (bp);
\draw[particle2] (ap) sin  (dp);
\draw[particle] (ap) sin  (cp);
\draw[particle2] (bp) sin  (dpp);
\draw[particle] (bp) sin  (cpp);
\fill[black] (a1) circle (.05cm);
\fill[black] (b2) circle (.05cm);
\draw (cp) circle (.06cm);
\draw (dp) circle (.06cm);
\draw (cpp) circle (.06cm);
\draw (dpp) circle (.06cm);
\end{tikzpicture} + \begin{tikzpicture}[node distance=0.6cm and 0.8cm]
\coordinate[label=below:$u_{t_2}$] (b2);
\coordinate[right=of b2,label=below:$y$,label=above right:$u$] (bp);
\coordinate[above=of bp,label=below left:$Q$,label=above:$y$,label=below right:$u$] (ap);
\coordinate[left=1.1cm of ap,label=above:$u_{t_1}$] (a1);
\coordinate[right=of ap](cc);
\coordinate[right=of bp](bb);
\coordinate[above=0.3cm of cc,label=right:$y_0$](cp);
\coordinate[below=0.1cm of cc,label=right:$u_0$](dp);
\coordinate[above=0.1cm of bb,label=right:$u_0$](cpp);
\coordinate[below=0.3cm of bb,label=right:$y_0$](dpp);
\draw[particle] (a1) -- (ap);
\draw[particle] (b2) -- (bp);
\draw[gluon] (ap) --  (bp);
\draw[particle2] (ap) sin  (dp);
\draw[particle] (ap) sin  (cp);
\draw[particle2] (bp) sin  (dpp);
\draw[particle] (bp) sin  (cpp);
\fill[black] (a1) circle (.05cm);
\fill[black] (b2) circle (.05cm);
\draw (cp) circle (.06cm);
\draw (dp) circle (.06cm);
\draw (cpp) circle (.06cm);
\draw (dpp) circle (.06cm);
\end{tikzpicture}.
\end{align}
Here, there is first the unconnected diagram, which represents the separate averages of $u_1$ and $u_2$. The remaining diagrams represent then the covariance of $u$ with itself at different times. Each of these uses two vertices with four legs, connected through either a $Q$ or an $I$ noise propagator. Note that we cannot use the 3-legged vertices because it connects at later time to only $x$ or $y$. Also, more diagrams of the above type cannot be generated with mixed initial conditions (with $x_0$ and $y_0$ on one diagram) because that would demand changing the noise flavor, which cannot be done.

Writing out the diagrams in terms of the Green functions gives the following for the $u-u$ autocovariance function,
\begin{align}
\nonumber & C_{uu} \equiv \la u(t_1) u(t_2)\ra - \la u(t_1)\ra \la u(t_2)\ra \nonumber\\
&= \zeta^2 u_0^2 x_0^2 \int dt' G_u(t_1,t') G_u(t_2, t')G_y(t',t_0)^2G_u(t',t_0)^2 \nonumber \\
&+ \zeta^2 u_0^2 y_0^2 \int dt' G_u(t_1,t') G_u(t_2, t')G_y(t',t_0)^2G_u(t',t_0)^2.
\end{align}
Evaluating the integrals gives the $u-u$ autocovariance function to be,
\be
{C}_{uu} = \frac{\zeta^2 u_0^2 (x_0^2+y_0^2)}{2 \gamma_2} e^{-\gamma_1(t_1+t_2)} (1 - e^{-2 \gamma_2 {\rm min}(t_1, t_2)}),
\ee
where ${\rm min}(t_1, t_2)$ indicates the minimum of $t_1$ and $t_2$.

We next consider the autocovariance function for the $x$-variables at two different times, given by
\begin{align}
\nonumber & C_{xx} = \la x(t_1) x(t_2) \ra - \la x(t_1) \ra \la x(t_2) \ra  \\
& = \!\!\!\!\begin{tikzpicture}[node distance=0.6cm and 0.8cm]
\coordinate[label=below:$x_{t_2}$] (b2);
\coordinate[right=of b2,label=below:$u$] (bp);
\coordinate[above=of bp,label=below left:$I$,label=above:$u$] (ap);
\coordinate[left=1.1cm of ap,label=above:$x_{t_1}$] (a1);
\coordinate[right=of ap,label=above:$u_0$](cc);
\coordinate[right=of bp,label=below:$u_0$](bb);
\draw[particle] (a1) -- (ap);
\draw[particle] (b2) -- (bp);
\draw[gluon] (ap) --  (bp);
\draw[particle2] (ap) --  (cc);
\draw[particle2] (bp) --  (bb);
\fill[black] (a1) circle (.05cm);
\fill[black] (b2) circle (.05cm);
\draw (cc) circle (.06cm);
\draw (bb) circle (.06cm);
\end{tikzpicture}\nonumber +
\!\!\!\!\begin{tikzpicture}[node distance=0.6cm and 0.8cm]
\coordinate[label=below:$x_{t_2}$] (b2);
\coordinate[right=of b2,label=below:$x$,label=above right:$x$] (bp);
\coordinate[above=of bp,label=below left:$I$,label=above:$u$] (ap);
\coordinate[left=1.1cm of ap,label=above:$x_{t_1}$] (a1);
\coordinate[right=of ap,label=above:$u_0$](cc);
\coordinate[right=of bp](bb);
\coordinate[below=0.1cm of cc](dp);
\coordinate[above=0.1cm of bb,label=right:$x_0$](cpp);
\coordinate[below=0.3cm of bb,label=right:$x_0$](dpp);
\draw[particle] (a1) -- (ap);
\draw[particle] (b2) -- (bp);
\draw[gluon] (ap) --  (bp);
\draw[particle] (ap) sin  (cc);
\draw[particle2] (bp) sin  (dpp);
\draw[particle] (bp) sin  (cpp);
\fill[black] (a1) circle (.05cm);
\fill[black] (b2) circle (.05cm);
\draw (cc) circle (.06cm);
\draw (cpp) circle (.06cm);
\draw (dpp) circle (.06cm);
\end{tikzpicture}+\!\!\!\! \begin{tikzpicture}[node distance=0.6cm and 0.8cm]
\coordinate[label=below:$x_{t_2}$] (b2);
\coordinate[right=of b2,label=below:$u$] (bp);
\coordinate[above=of bp,label=below left:$I$,label=above:$x$,label=below right:$x$] (ap);
\coordinate[left=1.1cm of ap,label=above:$x_{t_1}$] (a1);
\coordinate[right=of ap](cc);
\coordinate[right=of bp,label=below:$u_0$](bb);
\coordinate[above=0.3cm of cc,label=right:$x_0$](cp);
\coordinate[below=0.1cm of cc,label=right:$x_0$](dp);
\draw[particle] (a1) -- (ap);
\draw[particle] (b2) -- (bp);
\draw[gluon] (ap) --  (bp);
\draw[particle2] (ap) sin  (dp);
\draw[particle] (ap) sin  (cp);
\draw[particle2] (bp) sin  (bb);
\fill[black] (a1) circle (.05cm);
\fill[black] (b2) circle (.05cm);
\draw (cp) circle (.06cm);
\draw (dp) circle (.06cm);
\draw (bb) circle (.06cm);
\end{tikzpicture}\\
&+ \begin{tikzpicture}[node distance=0.6cm and 0.8cm]
\coordinate[label=below:$x_{t_2}$] (b2);
\coordinate[right=of b2,label=below:$x$,label=above right:$x$] (bp);
\coordinate[above=of bp,label=below left:$I$,label=above:$x$,label=below right:$x$] (ap);
\coordinate[left=1.1cm of ap,label=above:$x_{t_1}$] (a1);
\coordinate[right=of ap](cc);
\coordinate[right=of bp](bb);
\coordinate[above=0.3cm of cc,label=right:$x_0$](cp);
\coordinate[below=0.1cm of cc,label=right:$x_0$](dp);
\coordinate[above=0.1cm of bb,label=right:$x_0$](cpp);
\coordinate[below=0.3cm of bb,label=right:$x_0$](dpp);
\draw[particle] (a1) -- (ap);
\draw[particle] (b2) -- (bp);
\draw[gluon] (ap) --  (bp);
\draw[particle2] (ap) sin  (dp);
\draw[particle] (ap) sin  (cp);
\draw[particle2] (bp) sin  (dpp);
\draw[particle] (bp) sin  (cpp);
\fill[black] (a1) circle (.05cm);
\fill[black] (b2) circle (.05cm);
\draw (cp) circle (.06cm);
\draw (dp) circle (.06cm);
\draw (cpp) circle (.06cm);
\draw (dpp) circle (.06cm);
\end{tikzpicture} + \begin{tikzpicture}[node distance=0.6cm and 0.8cm]
\coordinate[label=below:$x_{t_2}$] (b2);
\coordinate[right=of b2,label=below:$y$,label=above right:$x$] (bp);
\coordinate[above=of bp,label=below left:$Q$,label=above:$y$,label=below right:$x$] (ap);
\coordinate[left=1.1cm of ap,label=above:$x_{t_1}$] (a1);
\coordinate[right=of ap](cc);
\coordinate[right=of bp](bb);
\coordinate[above=0.3cm of cc,label=right:$y_0$](cp);
\coordinate[below=0.1cm of cc,label=right:$x_0$](dp);
\coordinate[above=0.1cm of bb,label=right:$x_0$](cpp);
\coordinate[below=0.3cm of bb,label=right:$y_0$](dpp);
\draw[particle] (a1) -- (ap);
\draw[particle] (b2) -- (bp);
\draw[gluon] (ap) --  (bp);
\draw[particle2] (ap) sin  (dp);
\draw[particle] (ap) sin  (cp);
\draw[particle2] (bp) sin  (dpp);
\draw[particle] (bp) sin  (cpp);
\fill[black] (a1) circle (.05cm);
\fill[black] (b2) circle (.05cm);
\draw (cp) circle (.06cm);
\draw (dp) circle (.06cm);
\draw (cpp) circle (.06cm);
\draw (dpp) circle (.06cm);
\end{tikzpicture},
\end{align}
where we have used both three and four legged diagrams. By substituting the Green functions and evaluating time integrals, we get
\begin{eqnarray}
 C_{xx}&=&
\zeta^2 e^{-\gamma_2 (t_1+t_2)}\left[
\frac{u_0^2 \left( 1 - e^{-2(\gamma_1-\gamma_2) {\rm min} (t_1, t_2)} \right)}{2(\gamma_1-\gamma_2)}  \right. \nonumber \\
&-& \frac{2 u_0 x_0^2  \left(1 - e^{-\gamma_1 {\rm min} (t_1, t_2)} \right)}{\gamma_1} \nonumber \\
&+& \left. \frac{x_0^2(x_0^2+y_0^2)\left(1 - e^{-2 \gamma_2 {\rm min}(t_1, t_2)}\right) }{2\gamma_2}\right].
\label{cxx}
\end{eqnarray}
For the $y-y$ autocovariance function $C_{yy}$, it has the same form as $C_{xx}$ but replacing $x_0$ with $y_0$ and $y_0$ with $x_0$.

The cross-covariance function between different variables (including system variables and noise variables) can also be computed in the similar fashion. The cross-covariance function for $x$ and $y$ is given by,
\begin{align}
C_{xy}  =&  \begin{tikzpicture}[node distance=0.6cm and 0.8cm]
\coordinate[label=below:$y_{t_2}$] (b2);
\coordinate[right=of b2,label=below:$y$,label=above right:$x$] (bp);
\coordinate[above=of bp,label=above:$x$,label=below left:$I$,label=below right:$x$] (ap);
\coordinate[left=1.1cm of ap,label=above:$x_{t_1}$] (a1);
\coordinate[right=of ap](cc);
\coordinate[right=of bp](bb);
\coordinate[above=0.3cm of cc,label=right:$x_0$](cp);
\coordinate[below=0.1cm of cc,label=right:$x_0$](dp);
\coordinate[above=0.1cm of bb,label=right:$x_0$](cpp);
\coordinate[below=0.3cm of bb,label=right:$y_0$](dpp);
\draw[particle] (a1) -- (ap);
\draw[particle] (b2) -- (bp);
\draw[gluon] (ap) --  (bp);
\draw[particle2] (ap) sin  (dp);
\draw[particle] (ap) sin  (cp);
\draw[particle2] (bp) sin  (dpp);
\draw[particle] (bp) sin  (cpp);
\fill[black] (a1) circle (.05cm);
\fill[black] (b2) circle (.05cm);
\draw (cp) circle (.06cm);
\draw (dp) circle (.06cm);
\draw (cpp) circle (.06cm);
\draw (dpp) circle (.06cm);
\end{tikzpicture}
+\begin{tikzpicture}[node distance=0.6cm and 0.8cm]
\coordinate[label=below:$y_{t_2}$] (b2);
\coordinate[right=of b2,label=below:$y$,label=above right:$y$] (bp);
\coordinate[above=of bp,label=above:$y$,label=below left:$Q$,label=below right:$x$] (ap);
\coordinate[left=1.1cm of ap,label=above:$x_{t_1}$] (a1);
\coordinate[right=of ap](cc);
\coordinate[right=of bp](bb);
\coordinate[above=0.3cm of cc,label=right:$y_0$](cp);
\coordinate[below=0.1cm of cc,label=right:$x_0$](dp);
\coordinate[above=0.1cm of bb,label=right:$y_0$](cpp);
\coordinate[below=0.3cm of bb,label=right:$y_0$](dpp);
\draw[particle] (a1) -- (ap);
\draw[particle] (b2) -- (bp);
\draw[gluon] (ap) --  (bp);
\draw[particle2] (ap) sin  (dp);
\draw[particle] (ap) sin  (cp);
\draw[particle2] (bp) sin  (dpp);
\draw[particle] (bp) sin  (cpp);
\fill[black] (a1) circle (.05cm);
\fill[black] (b2) circle (.05cm);
\draw (cp) circle (.06cm);
\draw (dp) circle (.06cm);
\draw (cpp) circle (.06cm);
\draw (dpp) circle (.06cm);
\end{tikzpicture}  \nonumber \\
+&\!\!\!\!\begin{tikzpicture}[node distance=0.6cm and 0.8cm]
\coordinate[label=below:$y_{t_2}$] (b2);
\coordinate[right=of b2,label=below:$y$,label=above right:$x$] (bp);
\coordinate[above=of bp,label=above:$u$,label=below left:$I$] (ap);
\coordinate[left=1.1cm of ap,label=above:$x_{t_1}$] (a1);
\coordinate[right=of ap,label=above:$u_0$](cc);
\coordinate[right=of bp](bb);
\coordinate[below=0.1cm of cc](dp);
\coordinate[above=0.1cm of bb,label=right:$x_0$](cpp);
\coordinate[below=0.3cm of bb,label=right:$y_0$](dpp);
\draw[particle] (a1) -- (ap);
\draw[particle] (b2) -- (bp);
\draw[gluon] (ap) --  (bp);
\draw[particle] (ap) sin  (cc);
\draw[particle2] (bp) sin  (dpp);
\draw[particle] (bp) sin  (cpp);
\fill[black] (a1) circle (.05cm);
\fill[black] (b2) circle (.05cm);
\draw (cc) circle (.06cm);
\draw (cpp) circle (.06cm);
\draw (dpp) circle (.06cm);
\end{tikzpicture}+ \!\!\!\!\begin{tikzpicture}[node distance=0.6cm and 0.8cm]
\coordinate[label=below:$y_{t_2}$] (b2);
\coordinate[right=of b2,label=below:$u$] (bp);
\coordinate[above=of bp,label=above:$x$, label=below left:$Q$,label=below right:$y$] (ap);
\coordinate[left=1.1cm of ap,label=above:$x_{t_1}$] (a1);
\coordinate[right=of ap](cc);
\coordinate[right=of bp,label=below:$u_0$](bb);
\coordinate[above=0.3cm of cc,label=right:$x_0$](cp);
\coordinate[below=0.1cm of cc,label=right:$y_0$](dp);
\draw[particle] (a1) -- (ap);
\draw[particle] (b2) -- (bp);
\draw[gluon] (ap) --  (bp);
\draw[particle2] (ap) sin  (dp);
\draw[particle] (ap) sin  (cp);
\draw[particle2] (bp) sin  (bb);
\fill[black] (a1) circle (.05cm);
\fill[black] (b2) circle (.05cm);
\draw (cp) circle (.06cm);
\draw (dp) circle (.06cm);
\draw (bb) circle (.06cm);
\end{tikzpicture}
 \nonumber\\
 =& \zeta^2 e^{-\gamma_2(t_1 + t_2)}
 \left[\frac{x_0 y_0(x_0^2+y_0^2)
 \left(1 - e^{-2 \gamma_2 {\rm min}(t_1, t_2)}\right)}{2 \gamma_2}  \right. \nonumber \\
 -& \left. \frac{2 u_0 x_0 y_0 \left(1 - e^{- \gamma_1 {\rm min}(t_1, t_2)}\right)}{\gamma_1}  \right],
\end{align}
where $C_{xy} \equiv \la x(t_1) y(t_2) \ra - \la x(t_1) \ra\la y(t_2) \ra$. We also show the result of the cross-covariance between $u$ and $x$ which is,
\begin{align}
C_{ux}  =&  \begin{tikzpicture}[node distance=0.6cm and 0.8cm]
\coordinate[label=below:$x_{t_2}$] (b2);
\coordinate[right=of b2,label=below:$x$,label=above right:$x$] (bp);
\coordinate[above=of bp,label=above:$u$,label=below left:$I$,label=below right:$x$] (ap);
\coordinate[left=1.1cm of ap,label=above:$u_{t_1}$] (a1);
\coordinate[right=of ap](cc);
\coordinate[right=of bp](bb);
\coordinate[above=0.3cm of cc,label=right:$u_0$](cp);
\coordinate[below=0.1cm of cc,label=right:$x_0$](dp);
\coordinate[above=0.1cm of bb,label=right:$x_0$](cpp);
\coordinate[below=0.3cm of bb,label=right:$x_0$](dpp);
\draw[particle] (a1) -- (ap);
\draw[particle] (b2) -- (bp);
\draw[gluon] (ap) --  (bp);
\draw[particle2] (ap) sin  (dp);
\draw[particle] (ap) sin  (cp);
\draw[particle2] (bp) sin  (dpp);
\draw[particle] (bp) sin  (cpp);
\fill[black] (a1) circle (.05cm);
\fill[black] (b2) circle (.05cm);
\draw (cp) circle (.06cm);
\draw (dp) circle (.06cm);
\draw (cpp) circle (.06cm);
\draw (dpp) circle (.06cm);
\end{tikzpicture}
+\begin{tikzpicture}[node distance=0.6cm and 0.8cm]
\coordinate[label=below:$x_{t_2}$] (b2);
\coordinate[right=of b2,label=below:$y$,label=above right:$x$] (bp);
\coordinate[above=of bp,label=above:$y$,label=below left:$Q$,label=below right:$u$] (ap);
\coordinate[left=1.1cm of ap,label=above:$u_{t_1}$] (a1);
\coordinate[right=of ap](cc);
\coordinate[right=of bp](bb);
\coordinate[above=0.3cm of cc,label=right:$y_0$](cp);
\coordinate[below=0.1cm of cc,label=right:$u_0$](dp);
\coordinate[above=0.1cm of bb,label=right:$x_0$](cpp);
\coordinate[below=0.3cm of bb,label=right:$y_0$](dpp);
\draw[particle] (a1) -- (ap);
\draw[particle] (b2) -- (bp);
\draw[gluon] (ap) --  (bp);
\draw[particle2] (ap) sin  (dp);
\draw[particle] (ap) sin  (cp);
\draw[particle2] (bp) sin  (dpp);
\draw[particle] (bp) sin  (cpp);
\fill[black] (a1) circle (.05cm);
\fill[black] (b2) circle (.05cm);
\draw (cp) circle (.06cm);
\draw (dp) circle (.06cm);
\draw (cpp) circle (.06cm);
\draw (dpp) circle (.06cm);
\end{tikzpicture}  \nonumber \\
+&\!\!\!\!\begin{tikzpicture}[node distance=0.6cm and 0.8cm]
\coordinate[label=below:$x_{t_2}$] (b2);
\coordinate[right=of b2,label=below:$u$] (bp);
\coordinate[above=of bp,label=above:$u$, label=below left:$I$,label=below right:$x$] (ap);
\coordinate[left=1.1cm of ap,label=above:$u_{t_1}$] (a1);
\coordinate[right=of ap](cc);
\coordinate[right=of bp,label=below:$u_0$](bb);
\coordinate[above=0.3cm of cc,label=right:$u_0$](cp);
\coordinate[below=0.1cm of cc,label=right:$x_0$](dp);
\draw[particle] (a1) -- (ap);
\draw[particle] (b2) -- (bp);
\draw[gluon] (ap) --  (bp);
\draw[particle2] (ap) sin  (dp);
\draw[particle] (ap) sin  (cp);
\draw[particle2] (bp) sin  (bb);
\fill[black] (a1) circle (.05cm);
\fill[black] (b2) circle (.05cm);
\draw (cp) circle (.06cm);
\draw (dp) circle (.06cm);
\draw (bb) circle (.06cm);
\end{tikzpicture}
 \nonumber\\
 =& \zeta^2 e^{-\gamma_1 t_1 - \gamma_2 t_2}
 \left[\frac{u_0 x_0(x_0^2+y_0^2)
 \left(1 - e^{-2 \gamma_2 {\rm min}(t_1, t_2)}\right)}{2 \gamma_2}  \right. \nonumber \\
 -& \left. \frac{u_0^2  x_0 \left(1 - e^{- \gamma_1 {\rm min}(t_1, t_2)}\right)}{\gamma_1}  \right].
\end{align}
As before, the cross-covariance function for the $u$ and $y$ variables is in this same form, but changing $x_0 \rightarrow y_0$ and $y_0 \rightarrow x_0$.

\begin{figure*}
\includegraphics[width=17cm]{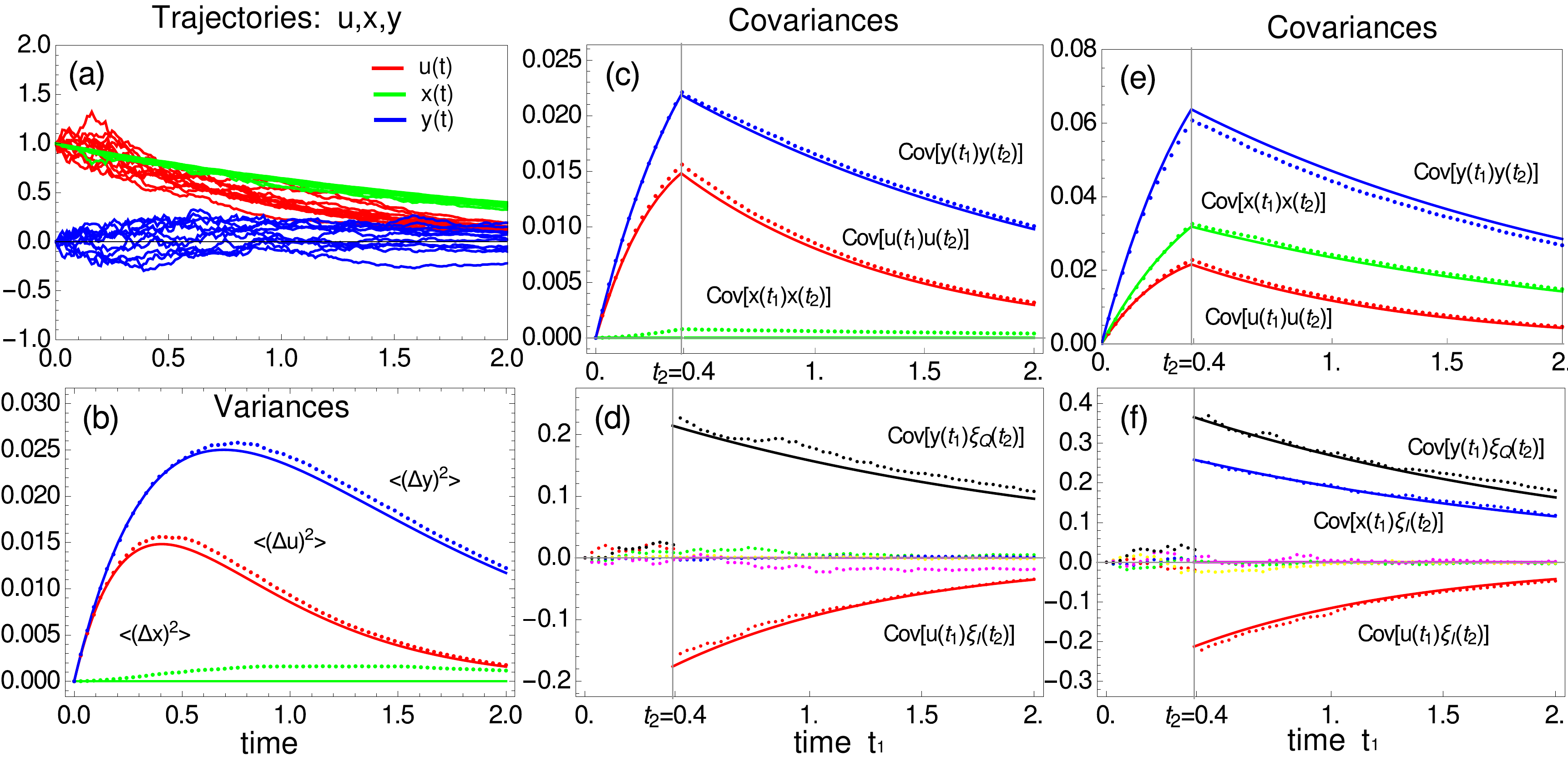}
\caption{The theoretically calculated variance and covariance functions (solid lines) are compared to their corresponding simulated data (dotted lines), for the system variables $u,x,y$ and the noise variables $\xi_I, \xi_Q$, given a fixed initial condition. Panel (a) shows sample trajectories generated with $u_0 = 1, x_0 = 1, y_0=0$, $\gamma_\phi = 0$, and $\eta = 0.2$, with $200$ time steps of size $0.01$ (time is plotted in units of $\gamma_1^{-1}$). To make a fair comparison, we simulate trajectories using the exact state update equations \eqref{eq-stateupdate} and generate the measurement outcomes at each time step with \eqref{eq-probalpha} and \eqref{rescale}. Panels (b,c,d) show the variances and autocovariances of qubit variables $u,x,y$, as well as their and cross-covariances with the noise variables $\xi_I, \xi_Q$ for the same set of data shown in panel (a). Other covariance functions are not shown that vanish for all time. Panels (e,f) show the autocovariances and cross-covariances for a different initial state $u_0 = 1/\sqrt{2}+1$, $x_0 =1/\sqrt{2}$, $y_0=0$, where in this case we see that all autocovariances of qubit variables are non-zero (there is a non-trivial $\text{Cov}[u(t_1)x(t_2)]$ not shown here). Note that we use the notation $\text{Cov}[a(t_1)b(t_2)] \equiv \la a(t_1) b(t_2) \ra - \la a(t_1) \ra \la b(t_2) \ra$ to explicitly show the time arguments. }
\label{fig-correlation}
\end{figure*}

Next, it is of interest to compute $\la r_i(t_1) \xi_k(t_2) \ra$, the correlation of the quantum trajectory variables with the noise variables. We note that the correlators of this form vanish if $t_1 \le t_2$ (the noise now can only affect the system in the future), so we consider only $t_1 > t_2$ in the equations below.
For the correlator between $u$ and $\xi_I$, there is one diagram at leading order,
\begin{align}\label{uxiI}
\la u(t_1) \xi_I(t_2) \ra &=\begin{tikzpicture}[node distance=0.6cm and 0.8cm]
\coordinate[label=below:$$] (b2);
\coordinate[right=of b2,label=left:$\xi_{I,t_2}$] (bp);
\coordinate[above=of bp,label=below left:$I$,label=above:$x$,label=below right:$u$] (ap);
\coordinate[left=1.1cm of ap,label=above:$u_{t_1}$] (a1);
\coordinate[right=of ap](cc);
\coordinate[right=of bp,label=below:$$](bb);
\coordinate[above=0.3cm of cc,label=right:$x_0$](cp);
\coordinate[below=0.1cm of cc,label=right:$u_0$](dp);
\draw[particle] (a1) -- (ap);
\draw[gluon] (ap) --  (bp);
\draw[particle2] (ap) sin  (dp);
\draw[particle] (ap) sin  (cp);
\fill[black] (a1) circle (.05cm);
\fill[black] (bp) circle (.05cm);
\draw (cp) circle (.06cm);
\draw (dp) circle (.06cm);
\end{tikzpicture} \nonumber  \\
&= - \zeta x_0 u_0 e^{-\gamma_1 t_1 - \gamma_2 t_2} \quad \text{for} \quad t_1 > t_2.
\end{align}
Similarly, the correlator $\la u(t_1) \xi_Q(t_2)\ra$ is of the same form, but with the replacement $x_0 \rightarrow y_0$, and using the corresponding vertices.
We then consider the correlation function between $x$ and the noise in the $I$-quadrature ($\xi_I$),
\begin{align}\label{xxiI}
\la x(t_1) \xi_I(t_2) \ra &=
\begin{tikzpicture}[node distance=0.6cm and 0.8cm]
\coordinate[label=below:$$] (b2);
\coordinate[right=of b2,label=left:$\xi_{I,t_2}$] (bp);
\coordinate[above=of bp,label=below left:$I$,label=above:$u$] (ap);
\coordinate[left=1.1cm of ap,label=above:$x_{t_1}$] (a1);
\coordinate[right=of ap,label=above:$u_0$](cc);
\coordinate[right=of bp,label=below:$$](bb);
\draw[particle] (a1) -- (ap);
\draw[gluon] (ap) --  (bp);
\draw[particle2] (ap) --  (cc);
\fill[black] (a1) circle (.05cm);
\fill[black] (bp) circle (.05cm);
\draw (cc) circle (.06cm);
\end{tikzpicture} + \
\begin{tikzpicture}[node distance=0.6cm and 0.8cm]
\coordinate[label=below:$$] (b2);
\coordinate[right=of b2,label=left:$\xi_{I,t_2}$] (bp);
\coordinate[above=of bp,label=below left:$I$,label=above:$x$,label=below right:$x$] (ap);
\coordinate[left=1.1cm of ap,label=above:$x_{t_1}$] (a1);
\coordinate[right=of ap](cc);
\coordinate[right=of bp,label=below:$$](bb);
\coordinate[above=0.3cm of cc,label=right:$x_0$](cp);
\coordinate[below=0.1cm of cc,label=right:$x_0$](dp);
\draw[particle] (a1) -- (ap);
\draw[gluon] (ap) --  (bp);
\draw[particle2] (ap) sin  (dp);
\draw[particle] (ap) sin  (cp);
\fill[black] (a1) circle (.05cm);
\fill[black] (bp) circle (.05cm);
\draw (cp) circle (.06cm);
\draw (dp) circle (.06cm);
\end{tikzpicture} \nonumber
 \\
&= \zeta u_0 e^{-\gamma_2 t_1 - (\gamma_1 - \gamma_2) t_2} - \zeta x_0^2 e^{-\gamma_2 (t_1 +t_2)},
\end{align}
for $t_1 > t_2$, and between $x$ and $\xi_Q$,
\begin{align}\label{xxiQ}
\la x(t_1) \xi_Q(t_2) \ra &=\begin{tikzpicture}[node distance=0.6cm and 0.8cm]
\coordinate[label=below:$$] (b2);
\coordinate[right=of b2,label=left:$\xi_{Q,t_2}$] (bp);
\coordinate[above=of bp,label=below left:$Q$,label=above:$x$,label=below right:$y$] (ap);
\coordinate[left=1.1cm of ap,label=above:$x_{t_1}$] (a1);
\coordinate[right=of ap](cc);
\coordinate[right=of bp,label=below:$$](bb);
\coordinate[above=0.3cm of cc,label=right:$x_0$](cp);
\coordinate[below=0.1cm of cc,label=right:$y_0$](dp);
\draw[particle] (a1) -- (ap);
\draw[gluon] (ap) --  (bp);
\draw[particle2] (ap) sin  (dp);
\draw[particle] (ap) sin  (cp);
\fill[black] (a1) circle (.05cm);
\fill[black] (bp) circle (.05cm);
\draw (cp) circle (.06cm);
\draw (dp) circle (.06cm);
\end{tikzpicture}\nonumber \\
&= - \zeta x_0 y_0 e^{-\gamma_2 (t_1+t_2)} \quad \text{for} \quad t_1 > t_2
\end{align}
Similarly, $\la y(t_1) \xi_Q(t_2) \ra $ is given by Eq.~(\ref{xxiI}) but with $x_0 \rightarrow y_0$, and $\la y(t_1) \xi_I(t_2)\ra$ is exactly given by Eq.~(\ref{xxiQ}).
Previously, we gave a general argument that higher order diagrams are suppressed by powers of $\zeta$. We illustrate this here by considering a higher order correction to the correlator $\la u(t_1) \xi_I(t_2) \ra$, where
\begin{align}
&\text{higher order} = \begin{tikzpicture}[node distance=0.6cm and 0.8cm]
\coordinate[label=below:$$] (b2);
\coordinate[right=of b2,label=left:$\xi_{I,t_2}$] (bp);
\coordinate[above=of bp,label=below left:$I$,label=above:$u$,label=below right:$x$] (ap);
\coordinate[left=1.1cm of ap,label=above:$u_{t_1}$] (a1);
\coordinate[right=of ap](cc);
\coordinate[right=of bp,label=below:$$](bb);
\coordinate[above=0.3cm of cc,label=below left:$I$,label=above:$u$](cp);
\coordinate[below=0.15cm of cc,label=below:$x$,label=above right:$x$](dp);
\coordinate[right=of cp](ccc);
\coordinate[right=of dp](cccc);
\coordinate[above=0.3cm of ccc,label=right:$u_0$](cpp);
\coordinate[below=0.1cm of ccc,label=right:$x_0$](dpp);
\coordinate[above=0.1cm of cccc,label=right:$x_0$](cppp);
\coordinate[below=0.3cm of cccc,label=right:$x_0$](dppp);
\draw[particle] (a1) -- (ap);
\draw[gluon] (ap) --  (bp);
\draw[gluon] (cp) --  (dp);
\draw[particle2] (ap) sin  (dp);
\draw[particle] (ap) sin  (cp);
\draw[particle2] (cp) sin  (cpp);
\draw[particle] (cp) sin  (dpp);
\draw[particle2] (dp) sin  (cppp);
\draw[particle] (dp) sin  (dppp);
\fill[black] (a1) circle (.05cm);
\fill[black] (bp) circle (.05cm);
\draw (cpp) circle (.06cm);
\draw (dpp) circle (.06cm);
\draw (cppp) circle (.06cm);
\draw (dppp) circle (.06cm);
\end{tikzpicture} \nonumber \\
&= -\zeta^3 x_0^3 u_0 \int dt' dt'' G_u(t_1 , t') G_u(t' , t'') G_x(t', t'')
\nonumber \\
& \times G_u(t'',0) G_x(t'',0)^3 \delta(t_2 -t').\nonumber \\
&= -\zeta^3 u_0 x_0^3  e^{-\gamma_1 t_1 - \gamma_2 t_2} \frac{1 - e^{-2 t_2 \gamma_2}}{2 \gamma_2}.
\end{align}
We see that this diagram is suppressed by at least $\eta x_0^2/2$ compared to the leading order diagram, with additional suppression for $2 \gamma_2 t_2 <1$.

\subsection{Comparison with numerical simulation}
The correlation functions presented in the previous subsection are computed from the tree-level diagrams, representing the first few orders of expansions in the small-noise approximation. As discussed in \cite{chantasri2015stochastic}, these approximate solutions can accurately describe the correlation functions of a system with low efficiency (small value of $\eta$), and in the short-time regime. We verify this by simulating $10^4$ qubit trajectories for $\eta = 0.2$ via a Monte Carlo method and compute their variances and covariances. The efficiency was chosen to be comparable to the experiments of Ref.~\cite{Campagne2015}.
As shown in Figure~\ref{fig-correlation}, the agreement between the approximate solutions and the numerical simulation is excellent. The variances in the system variables grows for some time to a maximum value and then drop to zero eventually (at the same time the qubit relaxes to its ground state), whereas the covariances between different variables decay as $t_1$ deviates from $t_2$. Moreover, the correlation functions between the system variables and the noise variables vanish whenever $t_1 \le t_2$, as predicted in the equations \eqref{uxiI}-\eqref{xxiQ}. We note that similar quality of agreement can still be seen with $\eta > 0.2$ and is acceptable up to $\eta \sim 0.5$. The approximate solutions at this expansion order fail to capture the system behavior as $\eta$ approaches $1$.

We note the appearance of certain \textit{magic} initial conditions, that cause a suppression of some noise correlators.  As shown in Fig.~(\ref{fig-correlation}a, \ref{fig-correlation}c), for the initial condition $u_0=1, x_0=1, y_0=0$, the subsequent diffusion of the $x$ variable is greatly suppressed to leading order in the expansion, so the $x-x$ covariance diagram vanishes.  We can see this explicitly by writing this covariance function (\ref{cxx}) in the special case where $\gamma_\phi=0, y_0=0$ as
\be
C_{xx} = \frac{\zeta^2}{\gamma_1}(u_0 - x_0^2)^2 e^{-\gamma_1 (t_1+t_2)/2} \left(1 - e^{-\gamma_1 {\rm min}(t_1, t_2)}\right).
\ee
It is clear that this correlator vanishes if $u_0 = 1, x_0=1$.  By changing the initial state to $u_0 = 1/\sqrt{2}+1$, $x_0 =1/\sqrt{2}$, $y_0=0$, the $x-x$ covariance function becomes nontrivial, as seen in Fig.~(\ref{fig-correlation}e).  Similar \textit{magic points} are $u_0=1, x_0=0, y_0=1$ for the $y-y$ covariance, and $u_0=2, x_0=0, y_0=0$ for the $u-u$ covariance.  Of course, in the ground state, $u_0=0, x_0=0, y_0=0$, all correlators vanish.  Higher order correlators may be calculated in a similar way.

\section{Conclusions}
We have considered the problem of a relaxing superconducting qubit, whose fluorescence is continuously monitored via a phase-preserving heterodyne measurement giving continuous quadrature measurement data. We have derived the probability distribution of short-time measurement results, as well as the measurement back-action on the state. This describes the relaxation of the qubit, but can also show counter-intuitive results where the conditioned measurement result causes the qubit to increase its energy. We then found stochastic differential equations describing the qubit dynamics. The solution of the most likely path between chosen boundary conditions, as well as the calculation of approximate correlation functions was carried out by reformulating the problem as a stochastic path integral, and applying the methods of that formalism. The same solution was found using quantum control theory.
We find good agreement between the results of those calculations and Monte Carlo numerical simulations of the trajectories and numerical averaging of the correlation functions. In particular, since in the experiments of Ref.~\cite{Campagne2015}, the efficiency of the measurement was about $0.24$, this improves the agreement, since the perturbation scheme is a small noise approximation, which is more accurate for smaller efficiency.

\section{Acknowledgments}
This work was supported by US Army Research Office Grants No. W911NF-09-0-01417 and No. W911NF-15-1-0496, by National Science Foundation grant DMR-1506081, by John Templeton Foundation grant ID 58558, and by Development and Promotion of Science and Technology Talents Project Thailand.     This work was partly supported by the EMERGENCES grant QUMOTEL of Ville de Paris. Thanks in particular to the COST Action MP1209 for supporting the Third Conference on Quantum Thermodynamics, in Porquerolles, France, where this work was begun.  We thank Mark Dykman, Alexander Korotkov, Kater Murch, and Alain Sarlette for discussions and helpful comments on the work.

\appendix

\section{Contextual Values}

In this appendix, we connect the POVM formalism with the question of what system operator the fluorescence is \textit{measuring}.
The contextual value formalism indicates what operators can be measured, based on the measurement that is being done.  Since the kind of measurement defines the \textit{context} of the experiment, this motivates the name \textit{contextual values}.  In our system, the measurement operator (\ref{mo}) defines the POVM elements $E_\alpha$ on the system, labeled by the continuous complex variable $\alpha$, written in the $|e\ra, |g\ra$ basis,
\be
E_\alpha = {\cal M}_\alpha^\dagger {\cal M}_\alpha =  \begin{pmatrix} 1-\epsilon( 1 - |\alpha|^2)  & \sqrt{\epsilon} \alpha   \\ \sqrt{\epsilon} \alpha^\ast & 1 \end{pmatrix} e^{-|\alpha|^2}.
\label{povmelements}
\ee
The contextual value approach \cite{cv1,cv2} allows us to \textit{target} given system observables ${ A}$ by defining contextual values $C_A(\alpha)$, such that operator $A$ may be constructed from the POVM elements,
\be
{ A} = \int \frac{d^2\alpha}{\pi} C_A(\alpha) E_\alpha.  \label{cvs}
\ee
The contextual values may be viewed as generalized eigenvalues of operator $A$, appropriate to the context of the measurement.

We stress that if the appropriate $C_A(\alpha)$ are chosen, then from repeated measurements from the same (generally unknown) state, we can reconstruct the (projective) averages of that operator $A$ from those weak measurements for any state, provided the construction exists.   That is,
\begin{eqnarray}
\la A \ra &=& {\rm Tr}[ \rho A] = \int \frac{ d^2\alpha}{\pi} C_A(\alpha) {\rm Tr}[\rho E_\alpha] \\
&=& \int \frac{d^2\alpha}{\pi} C_A(\alpha) P(\alpha) = \la C_A \ra_P,
\end{eqnarray}
where the last average $\la \ldots \ra_P$ is an average over the distribution of data in hand, $P(\alpha)$.
 While the inverse of this problem is not unique, and generally an infinite number of solutions exist, two typical solutions are the pseudo-inverse solution \cite{cv1,cv2} which works well when both $E(\alpha)$ and $A$ are diagonalizable in the same basis (which is not the case here), and a simple polynomial solution, which does work here.  It is easy to reconstruct the identity operator by setting $C_A(\alpha) = 1$.  We note that the two choices
\be
C_{\sigma_x}(\alpha) = \frac{2}{\sqrt{\epsilon}} {\rm Re}\, \alpha, \quad C_{\sigma_y}(\alpha) = \frac{-2}{\sqrt{\epsilon}} { \rm Im}\, \alpha,
\ee
are able to exactly construct the operators $\sigma_x$ and $\sigma_y$.  Thus, by repeated weak measurements of this form, we can extract the initial $x$ and $y$ components of any density matrix.  This is related to Eq.~(\ref{logp}), the distribution of the quadratures.
We can also find a contextual value construction for $\sigma_z$, although it takes many more samples of the distribution to accurately construct, because it involves an order $\epsilon$ shift of the distribution, rather than the order $\sqrt{\epsilon}$ of the first two Pauli matrices.  One can check by explicit calculation that the contextual value assignment
\be
C_{\sigma_z} = \frac{2}{\epsilon} (|\alpha|^2-1) - 1
\ee
is able to construct the operator $\sigma_z$.  This assignment may be interpreted as a shifted and rescaled average photon number from the measurement data.  Consequently, the fluorescence measurement results permit the construction of any qubit observable, and therefore can topographically construct an arbitrary initial state.

It is important to stress that the fluorescence measurement described in the main text is a continuous measurement, and not described by the procedure given above.  In particular, the state backaction is such that for repeated weak measurements, the state disturbance accumulates, and the system eventually relaxes to state $|g\ra$, and not to $\sigma_x$ or $\sigma_y$ eigenstates as one might expect.  This ultimately comes from the system operator in the interaction Hamiltonian being $\sigma_- = |g\ra\la e|$.

\section{Stochastic action  and most likely path for an arbitrary diffusive system }
This appendix generalizes the treatment given in the main text to an arbitrary quantum system being measured by an arbitrary number of continuous output signals.  We assume diffusive measurement dynamics, as well as the Markov approximation.  The stochastic action of the main text is generalized accordingly.  By specific choices of the measurement operators we introduce, the fluorescence measurement dynamics of the main text is recovered.

Consider first  the following stochastic master equation driven by a single Wiener process $W$ (here $H$ is an Hermitian operator, $L$ is a  measurement operator, and the detection efficiency is $\eta\in[0,1]$):
\begin{multline} \label{eq:Diff1D}
   \rho_{t+dt} -\rho_t=  d\rho = \left(-\imath[H,\rho]
    + \big( L\rho L^\dag  - \tfrac{1}{2} (L^\dag L\rho + \rho L^\dag L )\big) \right) dt \\+ \sqrt{\eta} \bigg( L \rho + \rho L^\dag - \tr{ L\rho + \rho L^\dag}\rho \bigg) dW,
\end{multline}
with the measured output signal $r_t =  \sqrt{\eta} \tr{(L+L^\dag)\rho} + \frac{dW}{dt}$.
It admits  the following discrete-time  formulation~\cite{rouchon2014models,rouchon2015efficient} where  $dt>0$ is  much smaller than the characteristic times associated to $H$ and $L$:
\begin{equation} \label{eq:Markov1D}
    \rho_{t+dt}  = \frac{ \tilde{M}_{r_t} \rho_t \tilde{M}_{r_t}^\dag + (1-\eta) \tilde{L}\rho_t \tilde{L}^\dag dt }
                              {\tr{\tilde{M}_{r_t} \rho_t \tilde{M}_{r_t}^\dag + (1-\eta) \tilde{L}\rho_t \tilde{L}^\dag dt}},
\end{equation}
with
\begin{align*}
  M_{r}&= I - (iH + L^\dag L/2) dt + \sqrt{\eta}\, r  L~dt
  \\
  R&= I+\left(-iH+\tfrac{L^\dag L}{2}\right) \left(iH+\tfrac{L^\dag L}{2}\right) dt^2
  \\
  \tilde{M}_r &= M_r  \left(\sqrt{R}\right)^{-1}
  \\
    \tilde{L} &= L \left(\sqrt{R}\right)^{-1}
    .
\end{align*}
Here the    probability to detect $r_t$ knowing $\rho_t$ is given by the following  density, depending  linearly  on  $\rho_t$:
\begin{multline}\label{eq:Proba1D}
   \PP{r_t \in[r,r+dr]~/~\rho_t} =
   \\ \tr{\tilde{M}_{r} \rho_t \tilde{M}_{r}^\dag + (1-\eta) \tilde{L}\rho_t \tilde{L}^\dag dt} \sqrt{\tfrac{ dt}{2\pi}}~ e^{-r^2 dt/2}~dr
   .
\end{multline}
The normalization corresponding to the right  multiplication of $M_r$ and $L$ by the inverse of $\sqrt{R}$ implies that
$$
\int_{-\infty}^{+\infty} \tfrac{\tr{M_{r} \rho_t M_{r}^\dag + (1-\eta) L\rho_t L^\dag dt} \sqrt{dt}}{\sqrt{2\pi}} e^{-r^2 dt/2}~dr  \equiv  1
.
$$
Notice that up-to  second order terms versus $dt$,  $\tilde{M}_r$ and $\tilde{L}$ coincide with $M_r$ and $L$.

This  discrete-time  formulation~\eqref{eq:Markov1D} converges in law (i.e. in distribution),  for $dt$ tending to $0^+$,  to the continuous time formulation~\eqref{eq:Diff1D}. This essentially results from the fact that    their  Markov generators, denoted below by  $A$ and $B$ respectively, coincide up to $O(\sqrt{dt})$ terms:
\begin{itemize}
  \item For any $C^2$ function $\rho \mapsto f(\rho)$, the expectation value of  $(f(\rho_{t+dt})-f(\rho_t))/dt$  knowing $\rho_t$ and calculated with~\eqref{eq:Diff1D} is, according to It${\hat {\rm o}}$ rules,
\begin{multline*}
  Af(\rho)= \dv{f}{\rho} \cdot \Big(-\imath[H,\rho]
    + \big( L\rho L^\dag  - \tfrac{1}{2} (L^\dag L\rho + \rho L^\dag L )\big) \Big)
    \\
    + \tfrac{\eta}{2} \dvv{f}{\rho}\cdot \Big(L\rho+\rho L^\dag - \tr{(L+L^\dag)\rho}\rho   ~,~ \\
  L\rho+\rho L^\dag - \tr{(L+L^\dag)\rho}\rho  \Big)
    .
\end{multline*}

 \item   the expectation value of  $(f(\rho_{t+dt})-f(\rho_t))/dt$  knowing $\rho_t$ and  based on~\eqref{eq:Markov1D}  is given by:
\begin{multline*}
   Bf(\rho_t) = \EE{\left.\frac{f(\rho_{t+dt})-f(\rho_t)}{dt}~\right/~\rho_t }
   \\= \int\limits_{-\infty}^{+\infty} \left( \tfrac{e^{-r^2 dt/2}  \tr{\tilde M_{r} \rho_t \tilde M_{r}^\dag + (1-\eta) \tilde L\rho_t \tilde L dt} }{\sqrt{2\pi~dt }}\right)
 \\ \left( f\left( \tfrac{ \tilde M_{r} \rho_t \tilde M_{r}^\dag + (1-\eta) \tilde L\rho_t \tilde L dt }
                              {\tr{\tilde M_{r} \rho_t \tilde M_{r}^\dag + (1-\eta) \tilde L\rho_t \tilde L dt}}\right) - f(\rho_t) \right)~dr
                              .
 \end{multline*}

\end{itemize}
Some  tedious computations yield $Bf(\rho_t) = Af(\rho_t) + O(\sqrt{dt})$.

Since for any $r$, we have
 \begin{multline*}
   \tilde M_{r} \rho \tilde M_{r}^\dag + (1-\eta) \tilde L\rho\tilde  L^\dag dt =
 \rho+ r \sqrt{\eta}  (L\rho+\rho L^\dag) dt
 \\
 +  \left( -i[H,\rho] +(1- \eta) L\rho L^\dag  - \tfrac{1}{2} (L^\dag L\rho + \rho L^\dag L ) \right) ~dt  + O(dt^2)
 \end{multline*}
 we get, up-to second order terms versus $dt$,
 \begin{multline}\label{eq:Markov1D_order1}
   \frac{\rho_{t+dt} - \rho_t }{dt} \approx   \mathcal{L}(\rho,r) \equiv
   -\imath[H,\rho] +  L\rho L^\dag  - \tfrac{1}{2} (L^\dag L\rho + \rho L^\dag L )  \\
  + \eta \big( \tr{L \rho L^\dag}\rho - L \rho L^\dag \big)\\+  r \sqrt{\eta} \Big( L \rho + \rho L^\dag - \tr{ L\rho + \rho L^\dag}\rho\Big)
 \end{multline}
 and
 \begin{multline}\label{eq:Proba1D_order1}
  \log\left(\tr{\tilde M_{r} \rho_t \tilde M_{r}^\dag + (1-\eta) \tilde L\rho_t \tilde L^\dag dt} e^{-r^2 dt/2}\right)
 \approx  \mathcal{F}(\rho,r) \\
 \equiv  \left(-r^2/2 + r \sqrt{\eta}  \tr{L\rho+\rho L^\dag}- \eta \tr{L\rho L^\dag} \right) ~dt
  .
\end{multline}
We have the following correspondences between the notations  of section II-A in~\cite{chantasri2013action} and  those used here:
\begin{itemize}
  \item $q$ becomes  here  $\rho$, an Hermitian operator of unit trace;
  \item   $p$ the adjoint state  becomes  here  $\xi$, an Hermitian operator;
  \item $p\cdot \dot q$ becomes here  $\tr{\xi \dot \rho}$.
\end{itemize}
The stochastic Hamiltonian defined by (3) in~\cite[section II A]{chantasri2013action}  reads  then
\begin{multline}\label{eq:StochasticHamiltonian}
\mathcal{H}(\xi,\rho,r) = \tr{\xi  \mathcal{L}(\rho,r)}
+ \mathcal{F}(\rho,r)\\
-\tr{\xi(\rho-\rho_I)}\delta(t) - \tr{\xi(\rho-\rho_F}\delta(t-T) .
\end{multline}
where, following the explanation  of~\cite[appendix A]{chantasri2013action}, the super-operator  $\mathcal{L}$ and $\mathcal{F}$ are given  by~\eqref{eq:Markov1D_order1} and~\eqref{eq:Proba1D_order1}.
Consequently, the first order stationary conditions (5a-5b-5c) of~\cite[section II B]{chantasri2013action} characterizing the most likely path become
  \begin{align*}
    &\dotex \rho   =  \mathcal{L}(\rho,r) \\
    &\dotex \xi   = -i[H,\xi] -  \left(L^\dag \xi L  - \tfrac{1}{2} (L ^\dag L \xi+ \xi L ^\dag L  )\right)
    \\& - \eta \big( \tr{L  \rho L ^\dag}\xi + (\tr{\xi\rho}-1) L^\dag L_ \nu  - L^\dag  \xi  L \big)
    \\& -   r  \sqrt{\eta } \Big( \xi L   +  L ^\dag \xi  - \tr{ L \rho + \rho L ^\dag}\xi - (\tr{\xi\rho}-1) (L +L^\dag)  \Big)
    \\
    &r = \sqrt{\eta} \Big( \tr{\xi (L  \rho + \rho L ^\dag)}  - \tr{ L \rho + \rho L ^\dag}(\tr{\xi \rho}-1) \Big)
    .
  \end{align*}

For arbitrary diffusive systems governed by
\begin{multline} \label{eq:DiffxD}
    d\rho = \left(-\imath[H,\rho] +\sum_\nu  L_\nu\rho L_\nu^\dag  - \tfrac{1}{2} (L_\nu^\dag L_\nu \rho + \rho L_\nu^\dag L_\nu ) \right) ~dt
    \\
    + \sqrt{\eta_\nu} \bigg( L_\nu \rho + \rho L_\nu ^\dag - \tr{ L_\nu\rho + \rho L_\nu^\dag}\rho \bigg) dW_\nu
\end{multline}
with  $m$ measured output signals,
$$
r_{\nu,t} =  \sqrt{\eta_\nu} \tr{(L_\nu+L_\nu^\dag)\rho} + \frac{dW_\nu}{dt} \text{ for }\nu=1,\ldots,m,
 $$
we have similarly the following discrete time formulation~\cite{rouchon2014models,rouchon2015efficient}:
$$
    \rho_{t+dt}  = \frac{ \tilde{M}_{\br_t} \rho_t \tilde{M}_{\br_t}^\dag + \sum_{\nu=1}^{m}(1-\eta_\nu) \tilde{L}_\nu\rho_t \tilde{L}_\nu^\dag dt }
                              {\tr{\tilde{M}_{\br_t} \rho_t \tilde{M}_{\br_t}^\dag + \sum_{\nu=1}^{m}(1-\eta_\nu) \tilde{L}_\nu\rho_t \tilde{L}_\nu^\dag dt}}
$$
with $\br_t=(r_{1,t},\ldots,r_{m,t})$
\begin{align*}
  M_{\br}&= I - (iH + \sum_{\nu=1}^{m}L^\dag L/2) dt + \sum_{\nu=1}^m \sqrt{\eta_\nu}\, r_{\nu} L_\nu~dt
  \\
  R&=I+\left(-iH+\sum_{\nu=1}^{m}\tfrac{L_\nu^\dag L_\nu}{2}\right)\left(iH+\sum_{\nu=1}^{m}\tfrac{L_\nu^\dag L_\nu}{2}\right) dt^2
  \\
  \tilde{M}_{\br} &= M_{\br} \left(\sqrt{R}\right)^{-1}
  \\
    \tilde{L}_\nu &= L_\nu   \left(\sqrt{R}\right)^{-1}
    .
\end{align*}
Here the    probability to detect $\br_t$ knowing $\rho_t$ is given by the following  density, depending  linearly  on  $\rho_t$:
\begin{multline*}
   \PP{\br_t \in\prod_{\nu=1}^{m}[r_\nu,r_\nu+dr_\nu]~\big/~\rho_t} =
   \\ \tr{\tilde{M}_{\br} \rho_t \tilde{M}_{\br}^\dag + \sum_{\nu=1}^{m}(1-\eta_\nu) \tilde{L}_\nu\rho_t \tilde{L}_\nu^\dag dt}
   \\
   \prod_{\nu=1}^{m}\sqrt{\tfrac{ dt}{2\pi}}~ e^{-r_\nu^2 dt/2}~dr_\nu
   .
\end{multline*}
Then, the stochastic Hamiltonian is given by~\eqref{eq:StochasticHamiltonian} with the following super-operators
\begin{align*}
 & \mathcal{L}(\rho,\br) = -\imath[H,\rho]
     + \sum_\nu \left(L_\nu\rho L_\nu ^\dag  - \tfrac{1}{2} (L_\nu ^\dag L_\nu \rho+ \rho L_\nu ^\dag L_\nu  )\right)
    \\
 & \qquad  + \sum_\nu  \left(\eta_\nu \big( \tr{L_\nu  \rho L_\nu ^\dag}\rho - L_\nu  \rho L_\nu ^\dag \big)\right)
 \\
  &\qquad + \sum_\nu \left(  r_{\nu} \sqrt{\eta_\nu} \Big( L_\nu  \rho + \rho L_\nu ^\dag - \tr{ L_\nu \rho + \rho L_\nu ^\dag}\rho\Big) \right).
    \\
&\mathcal{F}(\rho,\br)= \sum_\nu \Big( -r_\nu^2/2 \\
&+ r_\nu \sqrt{\eta_\nu}\tr{L_\nu \rho+\rho L_\nu^\dag } - \eta_\nu \tr{L_\nu \rho L_\nu ^\dag}\Big).
\end{align*}

For the fluorescence qubit,  the stochastic master equation~\eqref{eq-itoequations} corresponds to~\eqref{eq:DiffxD} with  the following set of operators
$$
H=0,~L_1=\sqrt{\tfrac{\gamma_1}{2}} \sigma_-,~L_2= iL_1 \text{ and } L_3=\sqrt{\tfrac{\gamma_\phi}{2}} \sigma_z,
$$
 the following detection efficiencies,   $\eta_1=\eta_2=\eta$ and  $\eta_3=0$,  and  only two  measurements $r_1=I$, $r_2=Q$, $r_3$  being  absent since $\eta_3=0$.
We have checked that $\dotex \rho = \mathcal{L}(\rho,\br)$  provides then, with  the modified Bloch sphere coordinates $(u,x,y)$,  system~\eqref{eq-mlpwitheta}.

\bibliography{fluorescencearXiv}

\end{document}